\documentclass[notitlepage,nofootinbib,preprintnumbers,amssymb]{revtex4-1}

\usepackage{graphicx} 
\usepackage{caption}
\usepackage{slashed} 
\usepackage{amsfonts}
\usepackage{enumerate} 
\usepackage{color}
\usepackage{amsmath} 
\usepackage{bm}
\usepackage[paperwidth=210mm,paperheight=280mm,centering,hmargin=2.2cm,vmargin=2.2cm]{geometry}

\begin{document}

\title{Multi-field formulation of gravitational particle production after inflation}

\author{Yuki Watanabe} \email{watanabe`at'resceu.s.u-tokyo.ac.jp}
\affiliation{Research Center for the Early Universe, University of Tokyo, Tokyo 113-0033, Japan}
\affiliation{Department of Physics, National Institute of Technology, Gunma College, Gunma 371-8530, Japan}
\author{Jonathan White} \email{jwhite`at'post.kek.jp}
\affiliation{Research Center for the Early Universe, University of Tokyo, Tokyo 113-0033, Japan}
\affiliation{Theory Center, KEK, Tsukuba 305-0801, Japan}

\begin{abstract}
We study multi-field inflation models that contain a non-trivial
field-space metric and a non-minimal coupling between the gravity and
inflaton sectors.  In such models it is known that even in the absence
of explicit interaction terms the inflaton sector can decay into matter
as a result of its non-minimal coupling to gravity, thereby reheating
the Universe gravitationally.  Using the Bogoliubov approach we evaluate
the gravitational decay rates of the inflaton fields into both scalars
and fermions, and analyse the reheating dynamics.  We also discuss how
the interpretation of the reheating dynamics differs in the so-called
Jordan and Einstein frames, highlighting that the calculation of the
Bogoliubov coefficients is independent of the frame in which one starts.
\end{abstract}

\preprint{RESCEU-5/15}
\maketitle

\section{Introduction}

An epoch of inflation in the very early Universe is now firmly supported
by recent observations of the cosmic microwave background (CMB)
\cite{Hinshaw:2012aka,Ade:2015tva,Planck:2015xua,Ade:2015lrj}.  These
observations suggest that the primordial curvature perturbation
generated during inflation is nearly Gaussian and adiabatic, with a
power spectrum that deviates from scale-invariance at the
5$\sigma$-level. They also suggest that the amplitude of tensor modes is
relatively small, i.e. that the energy scale of inflation is low.

With inflation widely accepted as a key part of the standard model of
cosmology, the question now turns to determining the exact nature of
inflation and how it might be embedded in some fundamental
high-energy-physics theory.  Inflation models containing non-minimal
gravitational coupling comprise one class of promising models.  As well
as being theoretically well motivated in the context of
high-energy-physics theories such as string theory, see
e.g. \cite{Copeland:1997ug}, their predictions also lie at the sweet
spot of current observational constraints \cite{Kaiser:2013sna}.
Examples include Starobinsky's original $R^2$ inflation (written in its
scalar-tensor form) \cite{Starobinsky:1980df}, Higgs inflation
\cite{Bezrukov:2007ep} and a whole class of so-called conformal
inflation models recently proposed by Kallosh {\it et al.}
\cite{Kallosh:2013yoa,*Kallosh:2013tua,*Kallosh:2013maa,*Kallosh:2013hoa,*Kallosh:2013lkr,*Kallosh:2013pby}.
Whilst most of these models are studied as single-field models, the
high-energy-physics theories that motivate them generically predict the
presence of multiple fields during inflation.  As such, it is important
to determine any possible signatures of multi-field effects in models
with non-minimal coupling
\cite{Kaiser:2012ak,White:2012ya,Greenwood:2012aj,White:2013ufa,Kallosh:2013daa,Schutz:2013fua}.

It has recently been demonstrated that in constraining specific models
of inflation with current CMB data, details of the reheating process
must be properly taken into account, even for single-field models
\cite{Dai:2014jja,Amin:2014eta,Martin:2014nya,Gong:2015qha}.  This is testimony to
the precision of current CMB data.  Moreover, in the context of
multi-field models of inflation, the primordial curvature perturbation
may continue to evolve during reheating, and this evolution must
therefore be tracked until an adiabatic limit is reached
\cite{Watanabe:2011sm,Leung:2012ve,Meyers:2013gua,Elliston:2014zea}.  It
is known that reheating can take place gravitationally in models with
non-minimal gravitational coupling; even if there are no explicit
interaction terms between the inflaton sector and matter,
gravitational particle production takes place as a result of the
non-minimal gravitational coupling
\cite{Starobinsky:1980df,Vilenkin:1985md,Bassett:1997az,Mijic:1986iv,Watanabe:2006ku,Faulkner:2006ub,Watanabe:2007tf,Gorbunov:2010bn,Watanabe:2010vy,Arbuzova:2011fu,Watanabe:2013lwa,Ema:2015dka}.
In light of the renewed interest in this class of models, in this paper
we revisit the theory of gravitational reheating after inflation and
present a multi-field formulation of gravitational particle production.
Whilst we focus on perturbative gravitational reheating, we nevertheless
employ the method of Bogoliubov transformations to determine the decay
rates.  As such, many of the features we discuss should also carry over
to the case of non-perturbative preheating.  Of course, the Bogoliubov
approach recovers the standard perturbative quantum-field-theory (QFT)
results when the appropriate limits are taken.

Throughout the analysis we try to pay particular attention to how the
interpretation of the reheating dynamics differs in the so-called Jordan
and Einstein frames.  During the oscillatory phase at the end of
inflation, the Hubble rate in the original Jordan frame contains an
oscillatory component, and it is this oscillatory component that gives
rise to particle production even in the absence of direct couplings
between the inflaton sector and the decay products.  The evolution of
the scale factor in the Einstein frame, on the other hand, is equivalent
to that of a matter-dominated universe, and can essentially be
neglected.  As such, the leading-order contribution to the gravitational
particle production is not a result of the oscillatory nature of the
Hubble rate.  In its place, however, one obtains explicit
gravitationally induced interaction terms between the inflaton sector
and ordinary matter, through which reheating proceeds.  Although the
interpretation in the two frames is different, we nevertheless find that
the calculation of the Bogoliubov coefficients is independent of the
frame in which we start; working in conformal coordinates and requiring
that the mode functions under consideration be canonically normalised
leads us to a common set of variables and form of action.

A technical complication that arises in the context of multi-field
models with non-minimal coupling is that even if one starts with a flat
field space in the Jordan frame -- i.e. a canonical, diagonal kinetic
term -- then one obtains a non-flat field space in the Einstein frame,
where evaluation of the inflaton dynamics is simpler
\cite{Kaiser:2010ps}.  We thus find it necessary to work in the mass
eigen-basis as defined with respect to the Einstein frame potential. As
a result, if either of the Jordan frame field-space metric or
non-minimal coupling are functions of some light spectator field, we
find that the gravitational decay rates generically become modulated,
giving rise to a modulated-reheating scenario
\cite{Dvali:2003em,Kofman:2003nx,Zaldarriaga:2003my,Suyama:2007bg}.

The rest of this paper is organised as follows: In Sec.\,\ref{model} we
begin by outlining the class of models under consideration and by
reviewing some of their key characteristics, with the review extending
into Appendix\,\ref{NM_rev}.  In Sec.\,\ref{rhd} we then analyse the
reheating dynamics.  We start, in Sec.~\ref{backdyn}, by looking at the
background dynamics of the oscillating inflaton fields at the end of
inflation, and in Sec.~\ref{QFT} and Sec.~\ref{Bogsec} we turn to the
reheating process itself, presenting the details of the Bogoliubov
calculation used to determine the decay rates.  Additional details
regarding the calculation of fermion production rates are included in
Appendix\,\ref{fermApp}.  Finally, Sec.\,\ref{sum} is devoted to summary
and conclusions.

\section{Multi-field models with non-minimal coupling}\label{model}

In this section we define more explicitly the class of models under
consideration and also discuss the relation between formulations made in
the Jordan and Einstein frames.

\subsection{Actions in the Jordan and Einstein frames}

The general class of models that we are considering take an action of
the form 
\begin{equation}\label{JFAc}
S = \int
d^4x\sqrt{-g}\left\{\frac{f(\phi)R}{2}-\frac{1}{2}h_{ab}g^{\mu\nu}\partial_\mu\phi^a\partial_\nu\phi^b
- V(\phi)\right\} + S_m,
\end{equation}
where $a,b = 1...n$ label $n$ scalar fields that are potentially all
non-minimally coupled to the Ricci scalar $R$ through the function
$f(\bm \phi)$.  $h_{ab}$ defines a non-flat field-space metric and $V$
is some general potential depending on all the fields.  We take the
matter part of the action to consist of bosons and fermions, namely
\begin{align}
S_m = \sum_i S_{\chi_i} + \sum_i S_{\psi_i},\qquad\mbox{where}\qquad\begin{array}{l}
S_{\chi_i} = \int
d^4x\sqrt{-g}\left\{-\frac{1}{2}g^{\mu\nu}\partial_\mu\chi_i\partial_\nu\chi_i
- U(\chi_i)\right\}, \\ \\S_{\psi_i} = -\int
d^4x\sqrt{-g}\left\{\overline{\psi}_i\overleftrightarrow{\slashed{D}}\psi_i
+ m_{\psi_i}\overline{\psi}_i\psi_i\right\}.\end{array}
\label{jordpsi}
\end{align} 
Here $\slashed{D}$ is given as $\slashed{D} = \gamma^\mu(x)D_\mu$, with
$D_\mu=\partial_\mu + \Gamma_\mu$ and
$\gamma^\mu(x)=e^\mu_\alpha\gamma^\alpha$, where $e^\mu_\alpha$ is the
tetrad defining local Lorentzian coordinates, $\gamma^\alpha$ are the
standard Dirac matrices satisfying\footnote{Note that we are working
with the signature $(- + + +)$.}
$\left\{\gamma^\alpha,\gamma^\beta\right\}=2\eta^{\alpha\beta}$, and the
spinor connection $\Gamma_\mu$ is defined as $\Gamma_\mu =
(1/2)\Sigma^{\alpha\beta}e_\alpha^\lambda\nabla_\mu e_{\beta\lambda}$,
where $\Sigma^{\alpha\beta}=\frac{1}{4}[\gamma^\alpha,\,\gamma^\beta]$.
We also have $\overline\psi_i = \psi_i^\dagger \beta$, where $\beta =
i\gamma^0$.  We have omitted the conformally invariant gauge fields, as
they do not play an important role in the perturbative reheating
considered in this paper.  See, however, \cite{Watanabe:2010vy} for a
discussion on the gauge trace anomaly and its importance in the
reheating process.

Matter in the above action is minimally coupled to gravity, and this
``frame'' is referred to as the Jordan frame.  However, on making the
conformal transformation $\tilde{g}_{\mu\nu}=\Omega^2g_{\mu\nu}$, with
$\Omega^2 = f(\bm \phi)/M_{\rm Pl}^2$, the action can be re-written as 
\begin{equation}\label{EFAct}
S = \int
d^4x\sqrt{-\tilde{g}}\left\{\frac{M_{\rm Pl}^2\tilde{R}}{2}-\frac{1}{2}S_{ab}\tilde{g}^{\mu\nu}\partial_\mu\phi^a\partial_\nu\phi^b
- \tilde{V}\right\} + S_{\tilde{m}},
\end{equation}
where\footnote{Note that, for example, $f_a$ denotes taking the derivative of $f$ with respect to
the $a$'th field.} 
\begin{equation}\label{metdef}
 S_{ab}=\frac{M_{\rm Pl}^2}{f}\left(h_{ab}+\frac{3f_af_b}{2f}\right),\qquad
\tilde{V}=\frac{M_{\rm Pl}^4V}{f^2}
\end{equation}
and the matter actions now take the form 
\begin{align}\label{EFMatt}
S_{\tilde{\chi}_i} &= \int
d^4x\sqrt{-\tilde{g}}\left\{-\frac{1}{2}\tilde{g}^{\mu\nu}\mathcal{D}_\mu\tilde{\chi}_i\mathcal{D}_\nu\tilde{\chi}_i
- \frac{U(\chi_i)}{\Omega^4}\right\},\\\label{EFMatt1}
S_{\tilde{\psi}_i} &= -\int
d^4x\sqrt{-\tilde{g}}\left\{\overline{\tilde{\psi}_i}\overleftrightarrow{\tilde{\slashed{D}}}\tilde{\psi}_i
+
\frac{m_\psi}{\Omega}\overline{\tilde{\psi}}_i\tilde{\psi}_i\right\}.
\end{align}
Here we have defined
\begin{align}
\tilde{\chi}_i = \frac{\chi_i}{\Omega},\quad \tilde{\psi}_i = \Omega^{-3/2}\psi_i,\quad
\mathcal{D}_\mu = \partial_\mu + \tilde{\chi}_i\partial_\mu(\ln\Omega)
 \quad\mbox{and}\quad
\tilde{\slashed{D}}
= \tilde{e}^\mu_\alpha\gamma^\alpha\left(\partial_\mu +
\Gamma_\mu\right),
\end{align}
where $\tilde{e}^\mu_\alpha = e^\mu_\alpha/\Omega$, which gives
$\tilde\gamma^\mu(x) = \gamma^\mu(x)/\Omega$, and the spinor connection
$\Gamma_\mu$ is conformally invariant (see, {\it e.g.}, footnote 4 of
\cite{Watanabe:2010vy}).  In this form, the fields $\phi^a$ are
minimally coupled to gravity and the gravity sector is of the standard
Einstein-Hilbert form.  The matter sector, however, becomes explicitly
coupled to the inflaton sector, and we must also be careful to take into
account the spacetime-dependent rescaling of units that results from the
conformal rescaling of the metric.

\subsection{Einstein's equations and the equations of motion}\label{EEEM}

Having defined our actions, let us briefly review the
gravitational equations of motion that they give rise to.  In this
section we will simply quote the main results.  Additional details
regarding these known results can be found in Appendix~\ref{NM_rev}.

Re-expressing \eqref{JFAc} in the form 
\begin{equation}\label{JFAc1}
 S=\int d^4x\sqrt{-g}\left\{\frac{f(\bm\phi)}{2}R +
		     \mathcal{L}^{(\phi)}+\mathcal{L}^{(m)}\right\},\qquad
 \mathcal{L}^{(\phi)} =
  -\frac{1}{2}h_{ab}g^{\mu\nu}\partial_\mu\phi^a\partial_\nu\phi^b - V(\bm{\phi}),
\end{equation}
where $\mathcal{L}^{(m)}$ contains the matter sector, and
minimising \eqref{JFAc1} with respect to $g^{\mu\nu}$ we get
\begin{equation}\label{eq:1}
 G_{\mu\nu}=
  \frac{1}{f}\left[T^{(\phi)}_{\mu\nu}+T^{(m)}_{\mu\nu}+\nabla_\mu\nabla_\nu
	       f-g_{\mu\nu}\Box
	       f\right],
\end{equation}
where
\begin{equation}\label{eq:2}
 T^{(\phi)}_{\mu\nu} =-\frac{2}{\sqrt{-g}}\frac{\delta\left(\sqrt{-g}\mathcal L^{(\phi)}\right)}{\delta g^{\mu\nu}}=h_{ab}\nabla_\mu\phi^a\nabla_\nu\phi^b -
  g_{\mu\nu}\left(\frac{1}{2}h_{ab}g^{\rho\sigma}\nabla_\rho\phi^a\nabla_\sigma\phi^b
	    + V\right).
\end{equation}
Similarly, varying the action with respect to the fields $\phi^a$ we get the
equations of motion 
\begin{equation}\label{eq:3}
 h_{ab}\Box\phi^b +
  \Gamma_{bc|a}g^{\mu\nu}\nabla_\mu\phi^b\nabla_\nu\phi^c - V_a + f_a R
  = 0,
\end{equation}
 where $\Gamma_{ab|c}=h_{cd}\Gamma^d_{ab}$ and $\Gamma^a_{bc}$ is the
 Christoffel connection associated with the field-space metric $h_{ab}$.  

Turning to the Einstein frame, we can similarly write the action
\eqref{EFAct} in the form
\begin{equation}
 S = \int d^4x\sqrt{-\tilde{g}}\left\{\frac{M_{\rm Pl}^2}{2}\tilde R + \tilde{\mathcal{L}}^{(\phi)}+\tilde{\mathcal{L}}^{(m)}\right\},\qquad
 \tilde{\mathcal{L}}^{(\phi)}=-\frac{1}{2}S_{ab}\tilde g^{\mu\nu}\partial_\mu\phi^a\partial_\nu\phi^b - \tilde{V},
\end{equation}
and we will return shortly to the relation between $\tilde{\mathcal L}^{(m)}$
and $\mathcal L^{(m)}$.  We then find the standard Einstein equations 
\begin{equation}\label{EFEE}
 \tilde{G}_{\mu\nu} = \frac{1}{M_{\rm Pl}^2}\left(\tilde{T}^{(\phi)}_{\mu\nu} +\tilde{T}^{(m)}_{\mu\nu}\right),
\end{equation}
where 
\begin{equation}
 \tilde T^{(\phi)}_{\mu\nu} =-\frac{2}{\sqrt{-\tilde g}}\frac{\delta\left(\sqrt{-\tilde g}\tilde{\mathcal L}^{(\phi)}\right)}{\delta \tilde g^{\mu\nu}}= S_{ab}\tilde\nabla_\mu\phi^a\tilde\nabla_\nu\phi^b -
  \tilde g_{\mu\nu}\left(\frac{1}{2}S_{ab}\tilde g^{\rho\sigma}\tilde\nabla_\rho\phi^a\tilde\nabla_\sigma\phi^b
	    + \tilde V\right).
\end{equation}
The equations of motion for the fields in the Einstein frame take the form 
\begin{equation}\label{EFFEM}
 -S_{ab}\tilde{\Box}\phi^b -
  ^{(S)}\Gamma_{bc|a}\tilde{g}^{\mu\nu}\tilde{\nabla}_\mu\phi^b\tilde{\nabla}_\nu\phi^c
  + \tilde{V}_{a} + \frac{\Omega_{a}}{\Omega}\tilde{T}^{(m)} = 0,
\end{equation}
where ${}^{(S)}\Gamma_{bc|a}=S_{ad}{}^{(S)}\Gamma^d_{bc}$ and
${}^{(S)}\Gamma^d_{bc}$ is the Christoffel connection associated with $S_{ab}$. 

Regarding the matter energy-momentum tensors, one can show (see Appendix
A for a review) that under certain conditions the following relations
hold:
\begin{equation}\label{genconteq}
 T^{(m)}_{\mu\nu} = \Omega^2 \tilde{T}_{\mu\nu}^{(m)},\qquad\nabla^\mu
  T^{(m)}_{\mu\nu}= 0\qquad\mbox{and}\qquad\tilde\nabla^\mu\tilde{T}^{(m)}_{\mu\nu}=-\frac{\Omega_\nu}{\Omega}\tilde{T}^{(m)}.
\end{equation}  

For future reference we note that the energy-momentum tensors associated
with $\chi_i$ and $\psi_i$ are given, respectively, as
\begin{align}\label{JFbosTmunu}
 T^{(\chi_i)}_{\mu\nu} &=
\nabla_\mu\chi_i\nabla_\nu\chi_i-g_{\mu\nu}\left(\frac{1}{2}g^{\rho\sigma}\nabla_\rho\chi_i\nabla_\sigma\chi_i
+ U(\chi_i)\right),\\\label{JFfermTmunu} T^{(\psi_i)}_{\mu\nu} &=
\frac{1}{2}\left(\overline\psi_i\gamma_{(\mu}(x)D_{\nu)}\psi_i -
D_{(\mu}\overline\psi_i\gamma_{\nu)}(x)\psi_i\right),
\end{align} 
where our symmetrisation with respect to the indices includes a
factor of a half.  In the Einstein frame we similarly have
\begin{align}\label{bosTmunu}
 \tilde T^{(\tilde\chi_i)}_{\mu\nu} &=
\mathcal D_\mu\tilde\chi_i\mathcal D_\nu\tilde\chi_i-\tilde g_{\mu\nu}\left(\frac{1}{2}\tilde g^{\rho\sigma}\mathcal D_\rho\tilde\chi_i\mathcal D_\sigma\tilde\chi_i
+  \frac{U(\tilde\chi_i)}{\Omega^4}\right),\\\label{fermTmunu} \tilde T^{(\tilde\psi_i)}_{\mu\nu} &=
\frac{1}{2}\left(\overline{\tilde\psi}_i\tilde\gamma_{(\mu}(x)D_{\nu)}\tilde\psi_i -
D_{(\mu}\overline{\tilde\psi}_i\tilde\gamma_{\nu)}(x)\tilde\psi_i\right).
\end{align}

\section{Reheating dynamics}\label{rhd}

Having described the general setup for our class of models and the
relation between the Jordan and Einstein frame formulations, in this
section we turn to the process of reheating.  We will begin by
considering the background dynamics of the inflaton fields after the end
of inflation, before then turning to the particle production process.
We also consider the effect of the produced particles on the dynamics of
the inflaton fields and how reheating ends.  At every step we try to
discuss how the interpretation of the reheating process differs in the
Jordan and Einstein frames.  For a review of reheating
after inflation and the techniques employed in this section, see
e.g. \cite{Kofman:1997yn,Allahverdi:2010xz,Amin:2014eta}.

\subsection{The oscillating phase}\label{backdyn}

In describing the dynamics of the inflaton fields after the end of
inflation we make the standard assumption that at background level our
Universe is described by a Friedmann-Lema\^{\i}tre-Robertson-Walker
(FLRW) metric.  Indeed, the homogeneity and isotropy of the Universe
should be guaranteed thanks to the preceding epoch of inflation.  If we
would like to write both the Jordan and Einstein frame metrics in FLRW
form, then the relation $\tilde{g}_{\mu\nu}=\Omega^2g_{\mu\nu}$ gives us
\begin{equation}\label{compBack}
d\tilde s^2 = -d\tilde t^2 +\tilde a^2(\tilde t)\delta _{ij}d\tilde
 x^id\tilde x^j = \Omega^2 ds^2 = \Omega^2\left(-dt^2 + a^2(t)\delta_{ij}dx^idx^j\right),
\end{equation}
where we recall that $\Omega^2 = f(\bm \phi)/M_{\rm Pl}^2$.  From the above
equation we then find the following relations:
\begin{align}
 \label{backRel}
\tilde a=\Omega
a,\quad
d{\tilde t}=\Omega dt,\quad
d\tilde{x}^i = dx^i \quad \mbox{and}\quad{\tilde H}
=\frac{1}{\Omega}
 \Big(
 H+\frac{\dot{\Omega}}{\Omega}
 \Big).
\end{align}
If we consider the epoch before reheating, when only the inflaton sector
is present, then the Friedmann equation and equations of motion for the
scalar fields in the Jordan frame are given, respectively, as
\begin{align}
3H^2 = \frac{1}{f}\left[\frac{1}{2}h_{ab}\dot\phi^a\dot\phi^b + V
 -3H\dot f\right]\qquad\mbox{and}\qquad \frac{D\dot\phi^a}{dt} + 3H\dot\phi^a +
 h^{ab}\left(V_b - f_bR\right)=0,\label{JFbg}
\end{align}
where $D\dot\phi^a/dt = \ddot\phi^a + \Gamma^a_{bc}\dot\phi^b\dot\phi^c$
and a dot denotes a derivative with respect to the Jordan frame
cosmic time $t$.  Similarly, in the Einstein frame we have 
\begin{align}\label{EFBGEQ}
3\tilde H^2 =
 \frac{1}{M_{\rm Pl}^2}\left[\frac{1}{2}S_{ab}\frac{d\phi^a}{d\tilde
 t}\frac{d\phi^b}{d\tilde t} + \tilde V\right]\qquad\mbox{and}\qquad
 \frac{\tilde D(d\phi^a/d\tilde t)}{d\tilde t} + 3\tilde
 H\frac{d\phi^a}{d\tilde t} +
 S^{ab}\tilde V_b=0,
\end{align}
where $\tilde D(d\phi^a/d\tilde t)/d\tilde t = d^2\phi^a/d\tilde{t}^2 +
{}^{(S)}\Gamma^a_{bc}(d\phi^b/d\tilde t)(d\phi^c/d\tilde t)$, $\tilde V
= M_{\rm Pl}^4 V/f^2$ and $\tilde t$ denotes the cosmic time in the Einstein
frame.

Given that the non-minimal coupling between the inflaton fields and
gravity is removed in transforming to the Einstein frame, it is much
more convenient to solve for the inflaton dynamics in this frame.  As
such, let us proceed by first solving for the dynamics in the Einstein
frame.  Do note, however, that it should be possible to solve directly
in the Jordan frame, see e.g. \cite{Arbuzova:2011fu}.

Given that in the Einstein frame the dynamics are determined by the
Einstein frame potential $\tilde{V}$ and the field-space curvature
associated with $S_{ab}$, we make the assumption that at the end of
inflation all of the inflaton fields begin to oscillate about the
minimum of $\tilde{V}$ at $\phi^a=\phi^a_{{\rm vev}}$, and can be
decomposed as $\phi^a = \phi^a_{{\rm vev}}+ \sigma^a$.  Requiring the
absence of a cosmological constant dictates that $\tilde V_{{\rm vev}} =
0$.  Combining this with the fact that $\tilde V_a |_{{\rm vev}} = 0$,
on expanding the inflaton part of the Einstein-frame action to second
order in $\sigma^a$ we get
\begin{equation}
S = \int d^4x\sqrt{-\tilde{g}}\left\{\frac{M_{\rm Pl}^2\tilde{R}}{2}-\frac{1}{2}S_{ab}|_{{\rm vev}}\tilde{g}^{\mu\nu}\partial_\mu\sigma^a\partial_\nu\sigma^b - \frac{1}{2}\tilde{V}_{ab}|_{{\rm vev}}\sigma^a\sigma^b\right\},
\end{equation}
where we have made the assumption that the potential can be well
approximated as being quadratic about its minimum.  In order to deal
with the non-diagonal nature of this action, we now introduce the mass
eigenstates of the Einstein-frame potential.  Namely, we take $\sigma^a
= e^a_A\alpha^A$, where
\begin{equation}
 \tilde V^a{}_b|_{{\rm vev}}e^b_A = m^2_{\hat A}e^a_A,
\end{equation}   
with $\tilde V^a{}_b|_{{\rm vev}} =
S^{ac}|_{{\rm vev}}\tilde V_{cb}|_{{\rm vev}}$
and $S_{ab}|_{{\rm vev}}e^a_Ae^b_B =
\delta_{AB}$.  In the above expression the hat on the index $A$ suppresses summation.   
The Einstein frame action then takes the form 
\begin{equation}
 S = \int d^4x\sqrt{-\tilde g}\left\{\frac{M_{\rm Pl}^2\tilde R}{2} + \frac{1}{2}\sum_A\left[\left(\frac{d\alpha^A}{d \tilde
	t}\right)^2 - m^2_{A}(\alpha^A)^2\right]\right\},
\end{equation}
so that the equations of motion for $\alpha^A$ are simply given as
\begin{equation}\label{alpha_eom}
\frac{d^2}{d\tilde{t}^2}\left(\tilde{a}^{3/2}\alpha^A\right) + \left[m_{\hat A}^2-\left(\frac{9}{4}\tilde{H}^2+\frac{3}{2}\frac{d\tilde{H}}{d\tilde{t}}\right)\right]\left(\tilde{a}^{3/2}\alpha^A\right) = 0.
\end{equation}
Note that in deriving the above results we have taken $\sigma^a$ and
hence $\alpha^A$ to be background quantities that only depend on time,
i.e. the decomposition $\phi^a = \phi^a_{\rm vev} + \sigma^a$ is not a
decomposition of $\phi^a$ into its classical background part and quantum
perturbation, but is simply an expansion of the classical part of
$\phi^a$ about $\phi^a_{\rm vev}$.  Also note that no terms involving
the field-space curvature appear in \eqref{alpha_eom}.  This is because
such terms would be second order in $\sigma^a$, taking the form
${}^{(S)}\Gamma^a_{bc}|_{\rm vev}(d\sigma^b/d\tilde t)(d\sigma^c/d\tilde
t)$.  As such, we see that models with very large field-space curvature,
such as that considered in Fig. 5 of \cite{Kaiser:2012ak}, lie beyond
the scope of our perturbative approach.

Making the assumption $m_A^2\gg
\tilde{H}^2,\, d\tilde{H}/d\tilde{t}$, i.e. that the timescale of the
field oscillations is much shorter than that of the background evolution
of the universe, we find the solutions
\begin{equation}\label{osEFH}
\alpha^A \simeq \frac{\alpha^A_0}{\tilde{a}^{3/2}}\cos[m_{\hat A} \tilde{t} + d_{\hat A}],
\end{equation}
where $d_A$ are constant phases.
The Einstein frame Friedmann equation then gives us   
\begin{equation}\label{EFHmb}
 \tilde H^2 = \frac{1}{6M^2_{{\rm
  Pl}}}\sum_A\left[\left(\frac{d\alpha^A}{d \tilde
	t}\right)^2 +m^2_{A}(\alpha^A)^2\right] = \sum_A\frac{(\alpha^A_0)^2m_A^2}{6M_{\rm Pl}^2\tilde a^3}\left(1 + \frac{3\tilde H}{2m_A}\sin(2(m_A\tilde t + d_A))+ \mathcal O(\tilde H^2/m_A^2)\right).
\end{equation}
As such,
we see that to leading order in $\tilde H/m_A$ the evolution of the
Einstein frame Hubble rate coincides with that of a matter-dominated
universe.  On calculating $d\tilde H/d\tilde t$ one finds 
\begin{equation}
 \frac{d\tilde H}{d\tilde t} \simeq -\frac{3}{2}\tilde H^2\left(1 - \frac{1}{\tilde H^2}\sum_A\frac{(\alpha^A_0)^2m_A^2}{6M_{\rm Pl}^2\tilde a^3}\cos(2(m_A\tilde t+ d_A))\right).
\end{equation}
The second term in the brackets represents an $\mathcal O(1)$ deviation
from the case of matter-domination, which can be seen by noting from
\eqref{EFHmb} that $(\alpha^A_0)^2m_A^2/(6M_{\rm Pl}^2\tilde a^3)\sim
\mathcal O(\tilde H^2)$.  However, the important result as far as we are
concerned is that $d \tilde H/d\tilde t\sim\tilde H^2\ll m_A^2$.

We are now interested in using these results to determine the background
evolution in the Jordan frame.  As is evident from \eqref{backRel}, in
order to do this we need expressions for $f$ and its derivatives, and
these can be obtained by expanding $f$ about $f_{{\rm vev}}$.  In doing
so, we make the assumption that by the end of reheating, when all fields
have decayed, $f_{{\rm vev}}=M_{\rm Pl}^2$.  We therefore have
\begin{equation}\label{fexpansion}
 f= M_{\rm Pl}^2\left(1 + \frac{f_a\sigma^a}{M_{\rm Pl}^2} +\frac{1}{2}\frac{f_{ab}\sigma^a\sigma^b}{M_{\rm Pl}^2} +...\right) =
  M_{\rm Pl}^2\left(1 + \frac{f_A\alpha^A}{M_{\rm Pl}^2} + \frac{1}{2}\frac{f_{AB}\alpha^A\alpha^B}{M_{\rm Pl}^2} +...\right),
\end{equation}
where $f_A = f_a e^a_A$.  Inserting this expansion into the last
relation in \eqref{backRel} and evaluating to leading order in
$\alpha^A$ and $\tilde H/m_A$ we get
\begin{equation}\label{JFH}
 H \simeq \tilde H\left(1+\frac{1}{\tilde H}\sum_A\frac{f_A}{2M_{{\rm
	      Pl}}^2}\frac{\alpha_0^A m_A}{\tilde a^{3/2}}\sin\left(m_A\tilde t + d_A\right)\right).
\end{equation}
If we assume that $f_A/M_{\rm Pl} \sim \mathcal O(1)$, and recall from \eqref{EFHmb} that $\alpha^A_0m_A/(M_{\rm Pl}\tilde a^{3/2})\sim
\mathcal O(\tilde H)$, we see that the evolution of the Hubble rate in the Jordan frame has an
oscillatory component that is not suppressed.  Note that to leading
order in $\tilde H/m_A$ the cosmic times as defined in the Jordan and
Einstein frames are interchangeable.  With this, we see that $\dot H$
picks up a term that is $\mathcal O(m_A \tilde H)$ (assuming
$f_A/M_{{\rm Pl}} \sim \mathcal O(1)$).  This is to be
compared with the case in the Einstein frame, where $d\tilde H/d\tilde t
\sim \mathcal O(\tilde H^2)$. 
   
\subsection{Perturbative QFT approach to reheating}\label{QFT}

Having discussed the background dynamics of the oscillating inflaton
fields at the end of inflation, there are essentially two ways in which
we can now consider reheating into ordinary matter.  The first follows
the standard perturbative QFT approach, and appears natural in the
Einstein frame.  The second method involves calculating Bogoliubov
coefficients in an approach based on QFT in a time-varying classical
background.  This second method appears natural in whichever frame we
begin, but the interpretation in each frame is somewhat different.  In
the case of perturbative reheating both methods are equally valid, and
the result is independent of the method used.

\subsubsection*{Decay rates}

In transforming to the Einstein frame, one consequence of the conformal
transformation is that we explicitly see the appearance of interaction
terms between the inflaton sector and ordinary matter.  These are
apparent in the factors of $\Omega$ that appear in $\mathcal{D}_\mu$,
$U/\Omega^4$ and $m_{\psi_i}/\Omega$ in \eqref{EFMatt} and
\eqref{EFMatt1}.  Using the expansion of $f$ given in \eqref{fexpansion},
and taking
\begin{equation}
U(\chi) = \frac{m_\chi^2\chi^2}{2},
\end{equation}
we find that the Einstein frame action contains the tri-linear interaction terms 
\begin{equation}\label{intlag}
\mathcal{L}^\chi_{int} =
 \frac{f_A\alpha^A}{4M_{\rm Pl}^2}\left(2m_\chi^2+m_{\hat{A}}^2\right)\tilde{\chi}^2\qquad\mbox{and}\qquad \mathcal{L}^\psi_{int}=\frac{f_A\alpha^A}{2M_{\rm Pl}^2}m_\psi\overline{\tilde{\psi}}\tilde{\psi},
\end{equation}
where we have integrated by parts and used the equations of motion for
$\alpha^A$ in deriving the first of these.\footnote{Note that as we have
used the background equations of motion for $\alpha^A$, the effective
interaction term is only valid in making tree-level calculations.  If we
wish to go beyond tree-level calculations, then we would have to use the
derivative interaction term directly.}  Here we neglect to consider
four-point interaction terms, as we know that such terms cannot allow
for complete reheating \cite{Kofman:1997yn}.  (See, however, Sec.~IV of \cite{Watanabe:2007tf}.)

Another key feature of the Einstein frame is that the scale factor is
evolving slowly, i.e.  $\tilde H^2,\,d\tilde H/d\tilde t \ll
m_A^2$, which allows us to neglect the expansion of the Universe.  As such, we
can use flat-space QFT calculations to determine the transition
amplitudes for $\alpha^A \rightarrow \tilde\chi\tilde\chi$ and
$\alpha^A\rightarrow \overline{\tilde\psi}\tilde\psi$ that result from
the interaction terms in \eqref{intlag}.  These amplitudes can in turn be used to
calculate the decay rates per unit time and volume of the oscillating
fields \cite{Watanabe:2006ku}:
\begin{align}\label{eq:9}
\tilde\Gamma_{\alpha^A\rightarrow \chi\chi} = \frac{\tilde{g}_{\chi A}^2}{8\pi m_{\hat{A}}}\left(1-\frac{4m_\chi^2}{m_{\hat{A}}^2}\right)^{1/2}\qquad\mbox{and}\qquad
\tilde\Gamma_{\alpha^A\rightarrow \overline{\psi}\psi} = \frac{\tilde{g}_{\psi A}^2 m_{\hat{A}}}{8\pi}\left(1-\frac{4m_\psi^2}{m_{\hat{A}}^2}\right)^{3/2},
\end{align}
where
\begin{equation}
\tilde{g}_{\chi A} = \frac{f_ae^a_A(m_{\hat{A}}^2+ 2m_\chi^2)}{4M_{\rm Pl}^2}\qquad\mbox{and}\qquad\tilde{g}_{\psi A} = \frac{f_ae^a_A m_\psi}{2M_{\rm Pl}^2}.
\end{equation}
In this approach we interpret the oscillating inflaton fields as a
condensate of zero-momentum particles that can decay into two scalars
or a fermion-anti-fermion pair.  Our reason for suggesting that
this approach seems ``natural'' in the Einstein frame is that it is in
this frame that the necessary interaction terms are explicit and that
the background evolution of the scale factor can be neglected.    

\subsubsection*{Dynamics including decay products}

Once the rate of decay becomes significant, namely
$\tilde\Gamma_A\sim\tilde H$ (see \eqref{GammaA} for the definition of $\tilde \Gamma_A$), we must take into account the effect
that the decay products have on the oscillating inflaton dynamics.
Remaining in the Einstein frame, we see from \eqref{EFFEM} that
the dynamics of the inflaton fields is sourced by the trace of the
matter energy-momentum tensor.  In Sec. \ref{backdyn} we
ignored this term, assuming that inflaton decay was initially negligible,
but now we must properly include it.  At the level of the action the
effect of matter fields on the dynamics of $\alpha^A$ is evident in the
explicit interaction terms, such as those given in \eqref{intlag}.  As
such, in the context of the perturbative QFT approach it is necessary to
calculate 1-loop corrections to the propagator of $\alpha^A$.  Invoking
the optical theorem, one finds that the effective equations of motion for
$\alpha^A$ take the form \cite{Kofman:1997yn}
\begin{equation}
\frac{d^2}{d\tilde{t}^2}\left(\tilde{a}^{3/2}\alpha^A\right) +
 \left[m_{\hat A}^2+im_{\hat A}\tilde\Gamma_{\hat A}-\left(\frac{9}{4}\tilde{H}^2+\frac{3}{2}\frac{d\tilde{H}}{d\tilde{t}}\right)\right]\left(\tilde{a}^{3/2}\alpha^A\right)
 = 0,
\end{equation}
where 
\begin{equation}\label{GammaA}
 \tilde\Gamma_{A} = \sum_i\tilde\Gamma_{\alpha^A\rightarrow
 \chi_i\chi_i} +
 \sum_j\tilde{\Gamma}_{\alpha^A\rightarrow\overline\psi_j\psi_j}.
\end{equation}
On inserting the zeroth-order solution for $\tilde H$, namely $\tilde{H}=2/3\tilde t$, the
last term in the square brackets vanishes.  If we also assume
$m_A\gg\tilde\Gamma_A$, then the
solutions to the above equations take the form
\begin{equation}\label{solsmod}
 \alpha^A = \frac{\alpha^A_0}{\tilde
		  a^{3/2}}\exp\left[-\frac{1}{2}\tilde\Gamma_{\hat A}\tilde{t}\right]\cos\left[m_{\hat A}\tilde
		  t + d_{\hat A} \right].
\end{equation}  
Comparing with \eqref{osEFH}, we see that there is an additional
exponential decay of the amplitude of the oscillations.  

Phenomenologically, the effect of inflaton decay is often modeled by
including an additional frictional term in the equations of motion for
$\alpha^A$ as follows \cite{Kofman:1997yn}: 
\begin{equation}\label{equiveq}
 \frac{d^2\alpha^A}{d\tilde{t}^2} + \left(3\tilde{H} + \tilde\Gamma_{\hat A}\right)\frac{d\alpha^A}{d\tilde{t}}
 + m_{\hat A}\alpha^A=0.
\end{equation} 
Indeed, under the assumptions $m_A\gg \tilde H, \Gamma_{A}$, one can see
that \eqref{solsmod} does satisfy this equation.  The advantage of using
this phenomenological equation is that it can be recast in a form that
is intuitive.  On multiplying through by $d\alpha^A/d\tilde t$ and
averaging over many cycles, it can be re-written as
\begin{equation}\label{alphacons}
 \frac{d\tilde{\rho}_A}{d\tilde{t}} + 3\tilde{H}\tilde{\rho}_A +
  \tilde\Gamma_{\hat A}\tilde{\rho}_A = 0, 
\end{equation}
where we have once again assumed $m_A\gg \tilde H,
\Gamma_A$ and 
\begin{equation}\label{alphaRho}
 \tilde{\rho}_A = \frac{1}{2}\left(\frac{d\alpha^A}{d\tilde t}\right)^2 +
  \frac{1}{2}m_{\hat A}^2(\alpha^A)^2.
\end{equation}
Summing over all $A$ we have 
\begin{equation}\label{alphaCont}
 \frac{d\tilde{\rho}_\alpha}{d\tilde{t}} + 3\tilde{H}\tilde{\rho}_\alpha +\tilde{\Gamma}_{\hat \alpha}\tilde{\rho}_\alpha = 0,
\end{equation}
where
\begin{equation}
 \tilde{\rho}_\alpha = \sum_A\tilde{\rho}_A\qquad \mbox{and} \qquad
  \tilde{\Gamma}_\alpha = \sum_A\frac{\tilde{\rho}_A}{\tilde{\rho}_\alpha}\tilde\Gamma_{A}. 
\end{equation}
We thus see that the energy density of the oscillating fields decays as a
result of the Hubble expansion and the decay into matter
particles, which is intuitively what we expect.  It is important to note, however, that this phenomenological
approach relies on the nature of the interaction terms considered and
the fact that the inflaton fields are oscillating in a quadratic
potential.  As such, the situation will be different in the more general case
\cite{Dolgov:1998wz}.
 
Assuming that the decay products quickly thermalise and can be modeled
as a relativistic fluid, we also have the equations
\begin{align}\label{boscons}
 &\frac{d\tilde{\rho}_{\chi_i}}{d\tilde{t}} + 4\tilde{H}\tilde{\rho}_{\chi_i} 
  -\sum_A\tilde{\Gamma}_{\alpha^A\rightarrow\chi_i\chi_i}\tilde{\rho}_A = 0,\\\label{fercons}
 &\frac{d\tilde{\rho}_{\psi_i}}{d\tilde{t}} + 4\tilde{H}\tilde{\rho}_{\psi_i} 
  -\sum_A\tilde{\Gamma}_{\alpha^A\rightarrow\overline\psi_i\psi_i}\tilde{\rho}_A = 0.
\end{align}
Combining with (\ref{alphacons}), one can see that the total energy
density is thus covariantly conserved, in agreement with \eqref{EFEE}.

\subsection{Bogoliubov approach to reheating}\label{Bogsec}

In the flat-space perturbative QFT approach discussed in the previous
subsection, we considered the oscillating inflaton fields as a
collection of massive zero-momentum particles decaying into matter
fields.  One of the limitations of this approach is that it can only be
applied in the perturbative regime, where interaction terms are small.
An alternative approach to calculating the decay rates is based on QFT
in a time-varying classical background.  Within this framework particle
production is a collective phenomenon, and the interaction terms do not
necessarily have to be small.  In the case that they are small, the
results of the previous subsection are recovered, but in the case that
the interaction terms are large it is possible to obtain resonant
particle production, or preheating -- see e.g. \cite{Kofman:1997yn}.  In
this paper we will focus on the perturbative regime and confirm
agreement with the perturbative QFT results given in the previous
subsection.  Much of our discussion, however, will also be relevant in
the preheating regime.

\subsubsection*{Jordan and Einstein frame interpretations}

In calculating particle production in a time-varying classical
background one is interested in solving for the mode functions of the
matter field under consideration,
\cite{Parker:1969au,Parker:1971pt,Zeldovich:1971mw,Starobinsky:1981vz},
and it is important that these are the mode functions associated with
canonically normalised fields.  As a consequence, we find that the
calculation becomes independent of the frame in which one starts,
leaving only a difference in interpretation.  To demonstrate this let us
consider the case of the bosonic field $\chi$.

Specialising to the case of an FLRW metric, the Jordan frame action for $\chi$ becomes
\begin{equation}
 S_\chi = \int dtd^3x a^3\frac{1}{2}\left[\dot\chi^2 -
				   \frac{1}{a^2}(\nabla\chi)^2 - m_\chi^2\chi^2\right],
\end{equation} 
where we have assumed $U(\chi) = m_\chi^2\chi^2/2$.  In order to bring
this into canonical form we use conformal time -- defined as $ad\eta =
dt$ -- and also introduce the re-scaled field $u = a\chi$, giving 
\begin{equation}\label{canChiAc}
 S_u = \int d\eta d^3x\frac{1}{2}\left[u^{\prime 2} -
				   (\nabla u)^2 - \left(a^2m_\chi^2-\frac{a^{\prime\prime}}{a}\right)u^2\right],
\end{equation} 
where a prime denotes differentiation with respect to conformal time.
Working in Fourier space this gives rise to the equations of motion for
the mode functions as
\begin{equation}\label{bosonmfe}
 u_k^{\prime\prime}+w_k^2u_k=0\qquad\mbox{with}\qquad w^2_k=k^2+a^2m_\chi^2-\frac{a^{\prime\prime}}{a},
\end{equation}
where $k = |\vec k|$.  As already discussed in Sec.~\ref{backdyn}, the scale factor in the
Jordan frame has a rapidly oscillating component, and it is the
resultant rapid time-variation of the effective mass of $u_k$ that gives
rise to particle production.  

If we now go to the Einstein frame, the action for $\tilde \chi$ takes
the form
\begin{equation}\label{EFchiAc}
 S_{\tilde\chi} = \int d\tilde td^3x \tilde
  a^3\frac{1}{2}\left[\left(\frac{d\tilde\chi}{d\tilde t}\right)^2 -
				   \frac{1}{\tilde a^2}(\nabla\tilde\chi)^2 - m_\chi^2\tilde\chi^2+\frac{f_A\alpha^A}{2M_{\rm Pl}^2}\left(2m_\chi^2+m_{\hat{A}}^2\right)\tilde{\chi}^2\right],
\end{equation}
where we have expanded $f$ to linear order in $\alpha^A$,
integrated by parts and used the equations of motion for $\alpha^A$ in
order to get the interaction terms.  In order to bring this into
canonical form we once again use conformal time (note that conformal
time is frame independent) and define the field
$\tilde u = \tilde a \tilde\chi$.  However, given that $\tilde\chi =
\chi/\Omega$, we see that 
\begin{equation}
\tilde u = \tilde a\tilde\chi = \frac{\tilde a}{\Omega}\chi = a\chi =
u. 
\end{equation}
As such, whichever frame we start in, the analysis becomes identical
once we transform to the canonically normalised variables.  Indeed, on
using \eqref{backRel} one finds
\begin{equation}
 \frac{a^{\prime\prime}}{a} = a^2\left(\dot H + 2H^2\right) =
  \frac{\tilde a^2 M_{{\rm Pl}}^2}{f}\left[\frac{f}{M_{{\rm
				      Pl}}^2}\left(\frac{d\tilde
					      H}{d\tilde t} + \tilde
					      H^2\right) -
				      \frac{1}{2M_{{\rm
				      Pl}}^2}\left(\frac{d^2f}{d\tilde
					      t^2}+3\tilde H
					      \frac{df}{d\tilde
					      t}\right) +
				      \frac{3}{4fM_{{\rm
				      Pl}}^2}\left(\frac{df}{d\tilde t}\right)^2\right].
\end{equation}
Expanding $f$ to first order in $\alpha^A$ and using the equations of
motion for $\alpha^A$ this reduces to 
\begin{equation}
 \frac{a^{\prime\prime}}{a} = \frac{\tilde a^{\prime\prime}}{\tilde a} +
  \frac{f_A\alpha^A}{2M_{{\rm Pl}}^2}\tilde a^2m_{\hat A}^2 + \mathcal O((\alpha^A)^2),
\end{equation} 
so that on substituting into \eqref{bosonmfe} we have 
\begin{equation}\label{eq:2}
 w_k^2\simeq k^2+\tilde{a}^2m_\chi^2 -\frac{\tilde a^{\prime\prime}}{\tilde a}- \frac{f_A\alpha^A}{M_{\rm Pl}^2}\tilde{a}^2m_\chi^2-\frac{1}{2}\frac{f_A\alpha^{A}}{M_{\rm Pl}^2}\tilde{a}^2m_{\hat{A}}^2,
\end{equation}
and this is in agreement with the frequency we would obtain from
\eqref{EFchiAc} on defining $\tilde u = \tilde a \tilde\chi$.  

The only
difference between the two frames, therefore, is the interpretation.  In
the Jordan frame the effective mass of the scalar field is oscillating
as a result of the oscillating scale factor, which is why we refer to
the process as gravitational reheating.  In the Einstein frame, however,
the scale factor is slowly varying and the oscillatory nature of the
effective mass of the scalar simply results from the explicit
interaction terms in the action.

Whilst above we have considered the case of a scalar field, one can
easily see that the same applies for fermions.  On using conformal time,
in the Jordan frame the
canonically normalised field is $\Psi = a^{3/2}\psi$.  In the Einstein
frame, on the other hand, we have 
\begin{equation}
\tilde\Psi = \tilde a^{3/2}\tilde\psi
= \left(\frac{\tilde a}{\Omega}\right)^{3/2}\psi = a^{3/2}\psi = \Psi. 
\end{equation}

\subsubsection*{The production of bosons}

We now turn to calculating the particle production rate for bosons, and
we follow very closely the analyses given in
\cite{Zeldovich:1971mw,Starobinsky:1981vz,Kofman:1997yn,Braden:2010wd}.
Let us start by expanding the quantum operator $u = a\chi$ in the standard
way as
\begin{equation}
 u = \int \frac{d^3k}{(2\pi)^{3/2}}\left[u_k a_{\vec k}e^{i\vec
				       k\cdot\vec x}+ u^\ast_k
				       a^\dagger_{\vec k}e^{-i\vec
				       k\cdot\vec x}\right],
\end{equation}
where $u_k$ satisfy \eqref{bosonmfe} and $a^\dagger_{\vec k}$ and
$a_{\vec k}$ are the creation and annihilation operators satisfying the
standard commutation relations.  

The Hamiltonian associated with the action \eqref{canChiAc} can then be
expanded as 
\begin{equation}
 {\rm H}_u = \frac{1}{2}\int d^3k\left[2E_k\left(2a^\dagger_{\vec k}a_{\vec k} +
				   \delta^{(3)}(0)\right) + F_ka_{\vec
 k}a_{-\vec k}+ F^\ast_ka^\dagger_{\vec
 k}a^\dagger_{-\vec k}\right],
\end{equation}
where 
\begin{equation}
 2E_k = |u_k^\prime|^2 + w_k^2|u_k|^2\qquad\mbox{and}\qquad F_k =
  (u_k^\prime)^2 + w_k^2u_k^2.
\end{equation}
It is possible to construct mode functions that satisfy $F_k =
0$ -- thus diagonalising the Hamiltonian -- as 
\begin{equation}
 u_k(\eta) =
  \frac{\alpha_k(\eta)}{\sqrt{2w_k(\eta)}}\exp\left[-i\int^\eta_{-\infty}d\eta^\prime
					    w_k(\eta^\prime)\right] + \frac{\beta_k(\eta)}{\sqrt{2w_k(\eta)}}\exp\left[i\int^\eta_{-\infty}d\eta^\prime
					    w_k(\eta^\prime)\right],
\end{equation}
where $\alpha_k(\eta)$ and $\beta_k(\eta)$ must satisfy the equations  
\begin{align}\label{bogeom}
 \alpha_k^\prime(\eta)&=\frac{w_k^\prime}{2w_k}\exp\left[2i\int^\eta_{-\infty}d\eta^\prime
 w_k(\eta^\prime)\right]\beta_k(\eta),\\
\beta_k^\prime(\eta)&=\frac{w_k^\prime}{2w_k}\exp\left[-2i\int^\eta_{-\infty}d\eta^\prime
 w_k(\eta^\prime)\right]\alpha_k(\eta).
\end{align}
We also require that $|\alpha_k|^2-|\beta_k|^2=1$ in order that the
canonical commutation relations for $u$ are satisfied.  One can confirm
that these mode functions also satisfy the equations of motion, and that
$E_k = w_k(1/2+|\beta_k(\eta)|^2)$.  If in the asymptotic past $w_k(\eta_0)$
is approximately constant, then the mode functions defined above are a linear
combination of the positive- and negative-frequency mode functions
associated with the Bunch-Davies vacuum.  If $\beta_k(\eta_0) = 0$ at
this time, then the mode functions do indeed coincide with those of the
Bunch-Davies vacuum, where $E_k$ is minimised.  As time evolves,
however, the evolution of $w_k(\eta)$ causes $\beta_k(\eta)$ to evolve
away from zero, meaning that $E_k$ is no longer minimised.  Given the
diagonal nature of the Hamiltonian, this effect can be
interpreted as particle production.

Let us now assume that at the initial time $\eta_0$ at which inflaton
oscillations commenced there are no $\chi$ particles present,
i.e. $\beta_k(\eta_0)=0$ and $\alpha_k(\eta_0) = 1$.  We then assume
that at times shortly after $\eta_0$ we are in the perturbative regime
where $\beta_k(\eta)\ll 1$ and $\alpha_k(\eta)-1\ll 1$.  This ensures
that we are in the same perturbative regime for which the QFT
calculations of Sec.~\ref{QFT} are applicable, where Bose condensate
effects are neglected.  Solving the above relations iteratively, we find
a solution for $\beta_k(\eta)$ as
\begin{align}\label{betaequation}
\beta_k(\eta)\simeq\int^\eta_{\eta_0}d\eta^\prime\frac{w_k^\prime}{2w_k}\exp\left[-2i\int^{\eta^\prime}_{-\infty}d\eta^{\prime\prime}
 w_k(\eta^{\prime\prime})\right].
\end{align}
Using \eqref{eq:2}, we find that $w_k^\prime$ and $w_k^2$ to leading order in $\alpha^A$ and $\tilde H/m_A$ are given as 
\begin{align}\label{simplew}
 w_k^\prime&\simeq-\frac{\tilde{a}^2}{2w_k}\frac{f_A\alpha^{A\prime}}{2M_{\rm Pl}^2}\left(m_{\hat A}^2+2m_\chi^2\right)+ \frac{\tilde a^3\tilde H m_\chi^2}{w_k},\\\label{simplew1}
w_k^2&\simeq k^2+\tilde{a}^2m_\chi^2.
\end{align}
Taking expressions to leading order in $\alpha^A$ ensures that we are
only considering the tri-linear interaction terms and perturbative regime
appropriate for comparison with the perturbative QFT calculations of
Sec.~\ref{QFT}.  In making order-of-magnitude estimates, we note that
from \eqref{EFHmb} one can deduce the order-of-magnitude relations
$\alpha^\prime/M_{\rm Pl} \sim \tilde a\tilde H$ and $m_{\hat
A}\alpha^A/M_{\rm Pl} \sim \tilde H$.  We have also assumed that
$f_A/M_{\rm Pl}\sim \mathcal O(1)$ and $k\sim \mathcal O(\tilde a m_A)$.
The first of these assumptions is compatible with expansion
\eqref{fexpansion} so long as we have $\alpha^A/M_{\rm Pl}\ll 1$, and
the second assumption comes from our expectation that modes with $k \sim
\mathcal O(\tilde a m_A)$ will be produced.  On substituting
\eqref{simplew}, \eqref{simplew1} and \eqref{osEFH} into
\eqref{betaequation} we notice that in general the integrand is highly
oscillatory in $\eta$.  As the second term in \eqref{simplew} is
non-oscillatory, we find that its contribution to $\beta_k(\eta)$
averages to zero.  The first term in \eqref{simplew}, however, is
oscillatory, which allows for the possibility of stationary points in
the total phase of the integrand.  We thus have
\begin{align}\label{eq:4}
 \beta_k(\eta)=\sum_A\frac{m_Af_A\alpha_0^A(2m_\chi^2+m_A^2)}{16M_{\rm
 Pl}^2i}\int^\eta_{\eta_0}d\eta^\prime\frac{\tilde{a}^{3/2}(\eta^\prime)}{k^2+\tilde{a}^2(\eta^\prime)m_\chi^2}\left\{\exp[im_A\psi_{k,1}^A(\eta^\prime)]-\exp\left[im_A\psi^A_{k,2}(\eta^\prime)\right]\right\},
\end{align}  
where 
\begin{align}\label{phase1}
 \psi_{k,1}^A(\eta^\prime)&=\int^{\eta^\prime}_{-\infty}d\eta^{\prime\prime}\left(\tilde
a - \frac{2w_k}{m_A}\right) + \frac{d_A}{m_{\hat A}},\\\label{phase2}
\psi_{k,2}^A(\eta^\prime)&=-\int^{\eta^\prime}_{-\infty}d\eta^{\prime\prime}\left(\tilde
a + \frac{2w_k}{m_A}\right) - \frac{d_A}{m_{\hat A}}.
\end{align}
The phase $\psi_{k,2}^A(\eta)$ does not have a stationary point for
physical values of $\tilde a$, so only the term involving
$\psi_{k,1}^A(\eta)$ contributes to $\beta_k(\eta)$.  Using the
stationary phase approximation we find
\begin{gather}\label{bosonbeta}
 \beta_k(\eta)=\sum_A\beta_k^A\exp\left[im_A\psi^A_{k,1}(\eta^A_{k})+is_k^A\pi/4\right]\Theta(\eta-\eta^A_{k})\Theta(\eta^A_k-\eta_0),\\\label{bosonbeta2}
\mbox{where}\qquad \beta^A_k = \frac{m_{\hat A}f_{\hat
A}\alpha_0^A(2m_\chi^2+m_{\hat A}^2)}{16M_{\rm Pl}^2i}\frac{4}{m_{\hat
A}^2}\sqrt{\frac{2\pi}{\tilde a(\eta_k^{\hat A}) m_{\hat A}|\psi^{{\hat
A}\prime\prime}_{k,1}(\eta^{\hat A}_{k})|}},
\end{gather}
$\eta^A_{k}$ is the time at which $d\psi^A_{k,1}/d\eta=0$ for some given
$k$ and $s_k^A$ is the sign of $\psi^{A\prime\prime}_{k,1}(\eta_k^A)$.
The phase is stationary when $2w_k = \tilde a m_A$, which to leading
order gives
\begin{align}\label{eq:5}
 \frac{k^2}{\tilde{a}^2(\eta^A_{k})} \simeq \frac{m_A^2}{4}\left(1-\frac{4m_\chi^2}{m_{\hat A}^2}\right).
\end{align}
Given that $k/\tilde a(\eta_k^A)$ coincides with the momentum of the
produced particle, this result coincides with our expectation from
kinematic considerations.  The second derivative of the phase is given
as 
\begin{equation}
 \psi^{A\prime\prime}_{k,1}(\eta_k^{\hat A}) \simeq \tilde a^2(\eta_k^{\hat A})\tilde
  H(\eta_k^{\hat A})\left(1-\frac{4m_\chi^2}{m_A^2}\right)+ \frac{2\tilde
	      a(\eta_k^A)}{m_{\hat A}^2}\sum_B\frac{f_B\alpha^{B\prime}(\eta_k^{\hat A})}{2M_{\rm
	      Pl}^2}\left(m_B^2 + 2m_\chi^2\right).
\end{equation}
The two step functions in the above expression
for $\beta_k(\eta)$ simply reflect the fact that a certain mode will
only have been excited if $\eta > \eta^A_k > \eta_0$.  Note that
$\eta^A_k$ is different for different $A$.

In looking to determine the production rate of $\chi$ particles let us
start by considering the continuity equation for the Einstein frame
energy-momentum tensor associated with $\tilde\chi$.  Under the
assumption of a FLRW background, we know that the expectation value of
the energy-momentum tensor can be expressed as $\langle 0|\tilde
T^{(\tilde\chi)\mu}{}_\nu|0\rangle={\rm diag}(-\tilde\rho_\chi,~\tilde
p_\chi,~\tilde p_\chi,~\tilde p_\chi)$, where $|0\rangle$ is the vacuum
state as defined with respect to $a_{\vec k}$.  The continuity equation
\eqref{genconteq} can then be written as
\begin{equation}\label{conteq}
 \frac{1}{\tilde a^4}\frac{d}{d\tilde t}(\tilde a^4\tilde\rho_\chi) + \tilde H(-\tilde\rho_\chi + 3\tilde
  p_\chi) = \frac{1}{2f}\frac{df}{d\tilde t}\left(-\tilde\rho_\chi +
					     3\tilde p_\chi\right).
\end{equation}
In the absence of particle production, i.e. in the absence of the
interaction terms given in \eqref{intlag}, the right-hand side of this
equation would be vanishing.  As such, in order to determine the
particle production rate, we wish to evaluate the right-hand side of
\eqref{conteq}.  

Using \eqref{bosTmunu} to determine $\tilde T^{(\tilde\chi)\mu}{}_\nu$,
re-expressing the result in terms of the canonically normalised field $u$ and taking the vacuum
expectation value, we find
\begin{align}\nonumber
 \tilde\rho_\chi &= \frac{1}{(2\pi)^3\tilde a^4}\int
 d^3k \left[w_k\left(\frac{1}{2}+|\beta_k|^2\right) - \mathcal
 H\Im\left(\alpha_k\beta^\ast_k e^{-2i\int w_k d\eta^\prime}\right)\right.\\
 &\qquad\hspace{2cm}\left.+\frac{1}{2w_k}(\mathcal H^\prime +
 2\mathcal H^2)\left(\frac{1}{2} + |\beta_k|^2+
 \Re\left(\alpha_k\beta_k^\ast e^{-2i\int w_k d\eta^\prime}\right)\right)
 \right],\\\nonumber -\tilde\rho_\chi+3\tilde p_\chi&=
 -\frac{1}{(2\pi)^3\tilde a^4}\int
 d^3k\left[2w_k\Re\left(\alpha_k\beta_k^\ast e^{-2i\int w_kd\eta^\prime}\right)
 + 2\mathcal H\Im\left(\alpha_k\beta_k^\ast e^{-2i\int
 w_kd\eta^\prime}\right)\right.\\
&\qquad\hspace{2cm}\left.+ \frac{1}{w_k}\left(\mathcal H^\prime
 +m_\chi^2 a^2\right)\left(\frac{1}{2}+|\beta_k|^2 +
 \Re\left(\alpha_k\beta_k^\ast e^{-2i\int w_kd\eta^\prime}\right)\right)\right],
\end{align}
where $\mathcal H = a^\prime /a$, $\Re(X)$ denotes the
real part of $X$ and $\Im(X)$ similarly the imaginary part.
Assuming that $\Im\left(\alpha_k\beta_k^\ast e^{-2i\int
w_kd\eta^\prime}\right)$ is of the same order of magnitude as
$\Re\left(\alpha_k\beta_k^\ast e^{-2i\int w_kd\eta^\prime}\right)$, keeping
terms only linear in $\beta_k$ and neglecting the vacuum density
contribution, we find that to lowest order in $\tilde{\mathcal H}/w_k$
the right-hand side of \eqref{conteq} is given as
\begin{align}\label{emtBog}
\frac{1}{2f}\frac{df}{d\tilde t}(-\tilde\rho_\chi+3\tilde p_\chi) &\simeq - 
\frac{1}{(2\pi)^3\tilde a^4}\frac{1}{2f}\frac{df}{d\tilde t}\int d^3k\frac{1}{w_k}\left[2w_k^2 +
						   m_\chi^2\tilde a^2\right]\Re\left(\alpha_k\beta_k^\ast e^{-2i\int
w_kd\eta^\prime}\right).
\end{align}
In general this quantity is highly oscillatory, and we are therefore
interested in finding its average over several oscillations,  i.e. over
a time-scale $T\sim \mathcal O(1/(\tilde a
m_A))$, where we have chosen to work in conformal time.  Taking $\alpha_k \simeq 1$ and
substituting the results \eqref{osEFH} and \eqref{bosonbeta} we have
\begin{align}\label{avRHS}
\left\langle\frac{1}{2f}\frac{df}{d\tilde t}(-\tilde\rho_\chi+3\tilde
 p_\chi)\right\rangle &= \frac{1}{2T}\int^{\eta + T}_{\eta -
 T}d\eta^\prime\sum_{A}\frac{4\pi}{(2\pi)^3\tilde a^4}\int dkk^2w_k \frac{\left[2w_k^2 +
						   m_\chi^2\tilde
 a^2\right]}{w_k^2}\frac{f_A\alpha_0^Am_A}{8M_{\rm Pl}^2\tilde
 a^{3/2}i}
 \\\nonumber&\hspace{3cm}\times 2i\Im\left[\beta^{\ast}_k
 \left(e^{im_A\psi^A_{k,1}(\eta^\prime)}-e^{im_A\psi^A_{k,2}(\eta^\prime)}\right)\right],
\end{align}  
where $\langle\rangle$ denotes taking the average over several
oscillations.  However, we can see that, due to the
oscillatory nature of the integrand, this average will only give a
non-zero result if $\eta$ coincides with a stationary point of
the phase of one of the terms in the integrand.  As there are no
stationary points of $\psi_{k,2}^A(\eta)$ for physical values of $\tilde
a$, the only terms giving a non-zero contribution are those containing
the phase $\psi_{k,1}^A(\eta)$, namely we find
\begin{align}
 \left\langle\frac{1}{2f}\frac{df}{d\tilde t}(-\tilde\rho_\chi+3\tilde
 p_\chi)\right\rangle &=\int dk\sum_{A}\delta(\eta-\eta_k^A)\int^{\eta_k^A + T}_{\eta_k^A -
 T}d\eta^\prime\frac{4\pi}{(2\pi)^3\tilde a^4}k^2 w_k\frac{\left[2w_k^2 +
						   m_\chi^2\tilde
 a^2\right]}{w_k^2}\frac{f_A\alpha_0^Am_A}{8M_{\rm Pl}^2\tilde
 a^{3/2}i}\\\nonumber&\hspace{5cm}\times 2i\Im\left[\beta^{\ast}_k
 e^{im_A\psi^A_{k,1}(\eta^\prime)}\right].
\end{align}
Seeing as the integral over $\eta^\prime$ is centred on the stationary
point for each $A$, we can take $T\rightarrow \infty$, as contributions
away from the stationary point will average to zero.  On making the
stationary phase approximation, and after a little manipulation we
eventually find 
\begin{align}\label{fullboson0}
\left\langle\frac{1}{2f}\frac{df}{d\tilde t}(-\tilde\rho_\chi+3\tilde p_\chi)\right\rangle &=
 \sum_{A,B}\frac{4\pi}{(2\pi)^3\tilde a^5(\eta)}\int dk \delta(\eta-\eta_k^A)
 k^2w_k(\eta_k^A)\beta_k^{A}\beta_k^{B\ast}
 \\\nonumber&\hspace{1.5cm}\times2\cos\left(m_A\psi_{k,1}^A(\eta_k^A)-m_B\psi_{k,1}^B(\eta_k^B)+(s_k^A-s_k^B)\pi/4\right)\Theta(\eta_k^A-\eta^B_{k})\Theta(\eta^B_k-\eta_0), 
\end{align}
where we have used \eqref{bosonbeta} and \eqref{bosonbeta2}.  If we
arrange that $m_B>m_A$ for $B>A$, meaning that
$\Theta(\eta_k^A-\eta_k^B) = 1$ only for $B>A$, then \eqref{fullboson0}
can be written as
\begin{align}\label{fullboson}
\left\langle\frac{1}{2f}\frac{df}{d\tilde t}(-\tilde\rho_\chi+3\tilde p_\chi)\right\rangle&=\sum_{A}\frac{4\pi}{(2\pi)^3\tilde a^5(\eta)}\int dk \delta(\eta-\eta_k^A)
 k^2w_k(\eta_k^A)|\beta_k^{A}|^2\\\nonumber
&\quad+\sum_{A,B>A}\frac{4\pi}{(2\pi)^3\tilde a^5(\eta)}\int dk \delta(\eta-\eta_k^A)
 k^2w_k(\eta_k^A)\beta_k^{A}\beta_k^{B\ast}\\\nonumber
&\hspace{1.5cm}\times 2\cos\left(m_A\psi_{k,1}^A(\eta_k^A)-m_B\psi_{k,1}^B(\eta_k^B)+(s_k^A-s_k^B)\pi/4\right)\Theta(\eta^B_k-\eta_0).
\end{align}
The delta function in $\eta$ can then be expressed as a delta function
in $k$ by using the fact that $\delta(\eta - \eta_k^A) =
|\psi_{k,1}^{A\prime\prime}(\eta_k^{\hat A})|\delta(\psi_{k,1}^{{\hat
A}\prime}(\eta))$ in combination with relation~\eqref{eq:5}, and we find
\begin{align}\label{deltaConv}
 \delta(\eta -\eta_k^A) =
 \frac{m_A^2}{4\mu_{\hat A}}|\psi_{k,1}^{{\hat A}\prime\prime}(\eta_k^{\hat A})|\delta(k-\tilde
a(\eta)\mu_{\hat A}),\qquad\mbox{where}\qquad \mu_A \simeq \frac{m_A}{2}\left(1-\frac{4m_\chi^2}{m_{\hat A}^2}\right)^{1/2}.
\end{align}
If we consider only the diagonal contributions to the double summation
in \eqref{fullboson}, i.e. the terms on the first line, and assume that
off-diagonal terms average to zero due to the cosine function, then we
find
\begin{align}
 \left\langle\frac{1}{2f}\frac{df}{d\tilde t}(-\tilde\rho_\chi+3\tilde p_\chi)\right\rangle
=\sum_A\frac{1}{\tilde{a}^3(\eta)}\frac{1}{256M_{\rm Pl}^4\pi}\frac{\left[m_Af_A\alpha^A_0(2m_\chi^2+m_A^2)\right]^2}{m_A}\left(1-\frac{4m_\chi^2}{m_A^2}\right)^{1/2},
\end{align} 
and by using the fact that
$\tilde{\rho}_{A}=m_{\hat A}^2(\alpha_0^A)^2/2\tilde{a}^3$, we can
then express this as
\begin{align}\label{bosresult}
 \left\langle\frac{1}{2f}\frac{df}{d\tilde t}(-\tilde\rho_\chi+3\tilde p_\chi)\right\rangle
=\sum_A\frac{1}{128M_{\rm Pl}^4\pi}\frac{\left[f_A(2m_\chi^2+m_A^2)\right]^2}{m_A}\left(1-\frac{4m_\chi^2}{m_A^2}\right)^{1/2}\tilde{\rho}_{A}\equiv\sum_A\tilde\Gamma_{\alpha^A\rightarrow\chi\chi}\tilde{\rho}_{A},
\end{align} 
i.e. we have found 
\begin{equation}
 \tilde\Gamma_{\alpha^A\rightarrow
  \chi\chi}=\frac{\left[f_A(2m_\chi^2+m_{\hat A}^2)\right]^2}{128\pi M_{\rm Pl}^4m_{\hat A}}\left(1-\frac{4m_\chi^2}{m_{\hat A}^2}\right)^{1/2},
\end{equation}
which is in agreement with \eqref{eq:9}.

\subsubsection*{The production of fermions}

In considering the case for fermions we follow closely the analyses of
\cite{DeWitt:1975ys,Peloso:2000hy,Nilles:2001fg,Chung:2011ck}.  As many
aspects of the calculation are similar to the bosonic case, we defer
details to Appendix \ref{fermApp}.  

From the fermionic action in \eqref{jordpsi} we obtain the Dirac equation
\begin{equation}
 \left(\gamma^\mu(x)D_\mu+m_\psi\right)\psi=0.
\end{equation}
Specialising to the case of a FLRW metric, and taking $e^0_a=\delta^0_a$ and
$e^i_{a}=\delta^i_a/a(t)$, we have
\begin{equation}
\gamma^0(x)=\gamma^0,\quad\quad\gamma^i(x)=\frac{\gamma^i}{a(t)},\qquad
 \Gamma_0 = 0 \quad\mbox{and}\quad\Gamma_i = \frac{\dot{a}}{2}\gamma^0\gamma^i.
\end{equation}
If we introduce conformal time, and also define
$\Psi=a^{3/2}\psi$, the Dirac equation can be written as
\begin{equation}
 \left(\gamma^a\partial_a+am_\psi\right)\Psi=0,
\end{equation}
where in this equation $\partial_0=\partial_\eta$.  Correspondingly, the
action and Hamiltonian can be written as 
\begin{equation}
 S_\Psi = -\int d\eta d^3x\overline\Psi\left(\gamma^a\partial_a +
				       am_\psi\right)\Psi\qquad\mbox{and}\qquad{\rm
 H}_\Psi = -\int d^3x\overline\Psi\gamma^0\partial_\eta\Psi = i\int d^3x\Psi^\dagger\partial_\eta\Psi.
\end{equation}
The space of solutions is endowed with a conserved scalar product, and
in the FLRW case it reduces to 
\begin{equation}
 \left(\Psi_1,\Psi_2\right)=\int d^3x\Psi_1^\dagger\Psi_2.
\end{equation}
Given one solution to the Dirac equation, $U_r(\vec k,x)$, one can show
that the charge conjugate $V_r(\vec k,x)=C\overline{U}^T_r(\vec k,x) =
\gamma^2U^\ast_r(\vec k,x)$, where $C=\gamma^2\beta$, is also a solution.  Note
that the subscript $r$ labels the spin.  We can then construct a basis of the
solution space out of $U_r(\vec k,x)$ and $V_r(\vec k,x)$, and further
require that the basis be orthonormal with respect to the above scalar product.
As such, a general solution can be decomposed as 
\begin{equation}\label{genspin}
\Psi(x) = \sum_r\int d^3k\left(a_r({ \vec k})U_r({\vec
				k},x)+b_r^\dagger({\vec k})V_r({\vec k},x)\right),
\end{equation}
where $a_r(\vec k)$ and
$b^\dagger_r(\vec k)$ now correspond to annihilation and creation
operators satisfying the anti-commutation relations
\begin{equation}
 \left\{a_r(\vec k),a_s^\dagger(\vec
  q)\right\}=\delta_{rs}\delta^{(3)}(\vec k-\vec
 q)\qquad\mbox{and}\qquad\left\{b_r(\vec k),b_s^\dagger(\vec
			  q)\right\}=\delta_{rs}\delta^{(3)}(\vec k-\vec
 q),
\end{equation}
with all other commutators vanishing.  We next decompose the solutions
$U_r({\vec k},x)$ as
\begin{equation}
 U_r({\vec k},x)=\frac{1}{(2\pi)^{3/2}}\left(\begin{array}{c}
				       u_{\mathcal A}(k,\eta)h_r(\hat{k})\\
					     ru_{\mathcal B}(k,\eta) h_r(\hat{k})
					    \end{array}\right)e^{i\vec
 k\cdot \vec x},
\end{equation}
where $\hat k = \vec k/k$ and $h_r(\hat{k})$ are the eigenvectors of the
helicity operator
\begin{equation}
 \hat{ k}\cdot\vec \sigma h_r(\hat{k})=rh_r(\hat{k}),\qquad r
  = \pm 1, 
\end{equation}
which are chosen to satisfy $h_r^\dagger(\hat
k)h_s(\hat k)=\delta_{rs}$.  For the choice of $h_r(\hat{k})$ made in
Appendix \ref{fermApp}, we then find
\begin{equation}
 V_r(\vec k,x)=\gamma^2U^\ast_r({\vec k},x)=\frac{e^{i\phi_{\hat k}}}{(2\pi)^{3/2}}\left(\begin{array}{c}
				       -u^\ast_{\mathcal B}(k,\eta)h_r(-\hat{k})\\
					     ru_{\mathcal A}^\ast(k,\eta) h_r(-\hat{k})
					    \end{array}\right)e^{-i\vec
 k\cdot \vec x}.
\end{equation}
Imposing the orthonormality conditions dictates that 
\begin{equation}
 |u_{\mathcal A}(k,\eta)|^2 + |u_{\mathcal B}(k,\eta)|^2 = 1,
\end{equation}  
and the Dirac equation now takes the form
\begin{equation}
i\partial_\eta
 \left(\begin{array}{c} u_{\mathcal A}(k,\eta)
  \\ u_{\mathcal B}(k,\eta)
 \end{array}\right)
=
\left(
\begin{array}{cc}
 am_\psi & k\\ k & -am_\psi
\end{array}\right)
\left(\begin{array}{c} u_{\mathcal A}(k,\eta)
  \\ u_{\mathcal B}(k,\eta)
 \end{array}\right),
\end{equation}
which can be decoupled to
\begin{equation}
 u_{\mathcal A,\mathcal B}^{\prime\prime}(k,\eta) =
  -\left[k^2+a^2m_\psi^2\pm i(am_\psi)^\prime\right]u_{\mathcal
  A,\mathcal B}(k,\eta),
\end{equation}
These two equations are now of the same form as Eq. \eqref{bosonmfe} for
the boson mode functions.  As such, the procedure from here onwards is
very similar to the bosonic case.  In analogy with the the bosonic case,
we expand $u_{\mathcal A}(k,\eta)$ and $u_{\mathcal B}(k,\eta)$ in terms
of positive and negative frequency functions as
\begin{align}
 u_{\mathcal A}(k,\eta) = \mathcal A_k(\eta)\sqrt{\frac{w_k+am_\psi}{2w_k}}e^{-i\int
 w_kd\eta^\prime}-\mathcal B_k(\eta)\sqrt{\frac{w_k-am_\psi}{2w_k}}e^{i\int w_kd\eta^\prime},\\
u_{\mathcal B}(k,\eta) = \mathcal A_k(\eta)\sqrt{\frac{w_k-am_\psi}{2w_k}}e^{-i\int
 w_kd\eta^\prime}+\mathcal B_k(\eta)\sqrt{\frac{w_k+am_\psi}{2w_k}}e^{i\int w_kd\eta^\prime},
\end{align}
where $w_k^2 = k^2+a^2m_\psi^2$.  With this decomposition we find that
$\mathcal A_k(\eta)$ and $\mathcal B_k(\eta)$ must satisfy the
normalisation condition $|\mathcal A_k(\eta)|^2 + |\mathcal B_k(\eta)|^2
= 1$ and the evolution equations
\begin{align}\label{fermBog}
\mathcal A_k^\prime(\eta) = -\frac{k(am_\psi)^\prime}{2w_k^2}e^{2i\int
 w_kd\eta^\prime}\mathcal B_k(\eta),\\\label{fermBog2}
\mathcal B_k^\prime(\eta) = \frac{k(am_\psi)^\prime}{2w_k^2}e^{-2i\int
 w_kd\eta^\prime}\mathcal A_k(\eta).
\end{align}
Assuming that at some time in the past $\mathcal B_k(\eta_0) = 0$, the
above mode functions then coincide with the flat-space mode functions
and the Hamiltonian is diagonal.  As $\mathcal B_k(\eta)$ evolve away
from zero, however, the Hamiltonian is no longer diagonal, instead
taking the form \cite{Peloso:2000hy,Nilles:2001fg,Chung:2011ck}
\begin{align}
 {\rm H}_\Psi &= \sum_r\int
  d^3k\left[w_k\left(|\mathcal A_k(\eta)|^2-|\mathcal B_k(\eta)|^2\right)\left(a_r^\dagger(\vec k)a_r(\vec
		k) - b_r(\vec k)b_r^\dagger(\vec k)\right)\right.\\\nonumber &\hspace{2.3cm}\left.-
  2\mathcal A_k(\eta)\mathcal B_k(\eta)w_ke^{-i\phi_{-\hat k}}b_r(-\vec k)a_r(\vec
  k)-2\mathcal A^\ast_k(\eta)\mathcal B^\ast_k(\eta)w_ke^{i\phi_{-\hat k}}a^\dagger(\vec k)b_r^\dagger(-\vec k)\right].
\end{align}
In order to diagonalise the Hamiltonian one can make a Bogoliubov 
transformation, defining
\begin{equation}
 \hat a_r(\vec k,\eta) = \mathcal A_k(\eta) a_r(\vec k) - \mathcal B_k^\ast(\eta) e^{i\phi_{-\hat k}}
  b_r^\dagger(-\vec k),\qquad \hat b^\dagger_r(\vec k,\eta) = \mathcal B_k(\eta)e^{-i\phi_{\hat k}}
  a_r(-\vec k) + \mathcal A_k^\ast(\eta) b_r^\dagger(k).
\end{equation} 
One then finds that the number operator associated with the new basis is
given as $\langle 0|\hat a_r^\dagger(\vec k,\eta) \hat
a_r(\vec k,\eta)|0\rangle = \langle 0|\hat b_r^\dagger(\vec k,\eta) \hat
b_r(\vec k,\eta)|0\rangle = |\mathcal B_k(\eta)|^2$.  We must therefore
determine $\mathcal B_k(\eta)$ if we wish to determine the number of
particles created.  Looking at \eqref{fermBog2} we see that the form of
the equation we need to solve for $\mathcal B_k(\eta)$ is almost
identical to that for $\beta_k(\eta)$ that we solved in the case of the
bosonic field.\footnote{Indeed, if we had considered a conformally
coupled field instead of a minimally coupled one, the evolution equation
for the Bogoliubov coefficients would be the same up to a factor of
$k/(ma)$ \cite{Starobinsky:1981vz}.}  As such, we defer details of the
calculation to Appendix~\ref{fermApp}, stating only the main results
here.

First, on using the stationary phase approximation we find that
$\mathcal B_k(\eta)$ is given as 
\begin{gather}\label{fermBeta}
 \mathcal B_k(\eta) = \sum_A\mathcal B_k^A\exp\left[im_A\psi^A_{k,1}(\eta^A_k)+is_k^A\pi/4\right]\Theta(\eta
  - \eta_k^A)\Theta(\eta_k^A-\eta_0), \\\label{BkA}\mathcal B_k^A = \frac{kf_{\hat A}m_{\hat A}\alpha_0^Am_\psi}{8M_{\rm Pl}^2iw_k^2(\eta_k^{\hat A})}\sqrt{\frac{2\pi\tilde a(\eta_k^{\hat A})}{m_{\hat A}|\psi^{{\hat A}\prime\prime}_{k,1}(\eta_k^{\hat A})|}},
\end{gather}
where $\psi_{k,1}^A(\eta)$ is still as defined in \eqref{phase1} but
with $w_k^2 = k^2 + a^2m_\psi^2$.  As in the bosonic case, $\eta_k^A$ is
the time at which $\psi_{k,1}^{A\prime}(\eta) = 0$ is satisfied,
i.e. the time at which the phase is stationary, and $s_k^A$ is the sign
of $\psi_{k,1}^{A\prime\prime}(\eta_k^{\hat A})$.  Then, as with the
scalar case, we wish to determine the quantity $(df/d\tilde t)\tilde T^{\tilde\psi}/(2f)$,
which corresponds to the right-hand side of the continuity equation for
the energy-momentum associated with $\tilde\psi$ in the Einstein frame,
i.e. corresponds to the particle production term.  Taking $\langle 0|\tilde
T^{(\tilde\psi)\mu}{}_\nu |0\rangle = {\rm
diag}(-\tilde\rho_\psi,~\tilde p_\psi,~\tilde p_\psi,~\tilde p_\psi)$,
we find that to first order in $\mathcal B_k(\eta)$ 
\begin{equation}\label{fermionRHS}
 \tilde T^{(\tilde\psi)} = -\tilde \rho_\psi + 3\tilde p_\psi \simeq
  -\frac{4}{(2\pi)^3\tilde a^4}\int d^3 k \frac{k\tilde a
  m_\psi}{w_k\Omega}\Re\left(\mathcal A_k(\eta)\mathcal B_k^\ast(\eta) e^{-2i\int^\eta_{-\infty} w_kd\eta^\prime}\right).
\end{equation} 
Proceeding in exactly the same way as for the bosonic field in the
previous subsection, we arrive at
\begin{align}\label{fermResult}
 \left\langle\frac{1}{2f}\frac{df}{d\tilde t}\left(-\tilde \rho_\psi + 3\tilde p_\psi\right)\right\rangle=\sum_A\tilde{\rho}_{A}\frac{(f_A)^2
 m_\psi^2 m_A}{32\pi M_{\rm Pl}^4}\left(1-\frac{4m_\psi^2}{m_A^2}\right)^{3/2},
\end{align} 
from which we deduce
\begin{equation}
 \tilde\Gamma_{\alpha^A\rightarrow\overline\psi\psi}=\frac{(f_A)^2
 m_\psi^2
 m_{\hat A}}{32\pi M_{\rm Pl}^4}\left(1-\frac{4m_\psi^2}{m_{\hat A}^2}\right)^{3/2}, 
\end{equation}
which is in agreement with \eqref{eq:9}.  

\subsubsection*{Energy-momentum tensors and their (non-)conservation}

In the analysis of the preceding two subsections we considered the
continuity equation for the matter energy-momentum tensor in the
Einstein frame.  Our reason for doing so was the transparent
interpretation: the non-conservation of the matter energy-momentum tensor in
the Einstein frame is a result of the explicit interaction terms that
give rise to particle production.  To close this section we consider
the continuity equations for the other energy-momentum tensors.

First let us consider the energy-momentum tensor of the oscillating
fields in the Einstein frame.  Combining the results of the previous
two subsections we have
\begin{equation}
\tilde\nabla_\mu \tilde T^{(m)\mu}{}_0 = -\sum_A \tilde\Gamma_A\tilde\rho_{A}. 
\end{equation}
As the total energy-momentum tensor must be conserved, this implies that
$\tilde\nabla_\mu\tilde
T^{(\phi)\mu}{}_0=\sum_A\tilde\Gamma_A\tilde\rho_{A}$, where
$\tilde T^{(\phi)\mu}{}_\nu = \sum_A{\rm diag}(-\tilde\rho_A, \tilde
p_A,\tilde p_A,\tilde p_A)$, with $\tilde\rho_A$ as given in
\eqref{alphaRho} and 
\begin{equation}
 \tilde p_A = \frac{1}{2}\left(\left(\frac{d\alpha^A}{d\tilde t}\right)^2 -
		      m_{\hat A}^2(\alpha^A)^2\right).
\end{equation}
Averaging over several oscillations we have $\langle \tilde p_A\rangle =
0$, so that we obtain the expected continuity equation
\eqref{alphaCont}.  As previously mentioned, the interpretation of this
standard result is intuitive -- the energy density of the oscillating
fields decays both as a result of the Hubble expansion and the decay
into matter.  Under the instant decay approximation, we assume that
reheating ends once the decay rate ``catches up'' with the Hubble
expansion, i.e. when $\tilde \Gamma_\alpha = 3\tilde H$.  This then
allows us to determine the reheating temperature in terms of $\tilde
\Gamma_\alpha$, and thus put constraints on model parameters such as
$f_A$ and $m_A$.

Next we consider the matter energy-momentum tensor in the
Jordan frame, which we know to be covariantly conserved -- recall
\eqref{genconteq}.  At first glance this property would seem to be
at odds with the fact that we have particle production.  However, in the
Jordan frame the particle production is
interpreted as being due to the oscillatory nature of the Hubble rate, and the term on the right-hand side
of \eqref{conteq}, for example, becomes part of the Hubble dilution term on the
left-hand side of the continuity equation in the Jordan frame.  Namely, using $\rho_\chi =
\Omega^4\tilde\rho_\chi$, $p_\chi = \Omega^4 \tilde p_\chi$ and the relations given
in \eqref{backRel}, \eqref{conteq} can be re-written as the standard continuity equation
\begin{equation}\label{JFBosCons}
 \frac{1}{a^4}\frac{d}{dt}(a^4\rho_\chi) + H(-\rho_\chi + 3p_\chi) = 0.
\end{equation}  
Note that whilst we have considered the bosonic field as an example, the
same is also true for any matter field.

Finally we consider the energy-momentum tensor for the oscillating
fields in the Jordan frame.  In the Jordan frame there is some ambiguity
as to how we might like to define the energy-momentum tensor of the
inflaton fields, and the relation between the Jordan and Einstein frame
inflaton energy-momentum tensors is not just a simple factor of
$\Omega^2$, as it is for the matter energy-momentum tensors.  As
commented in Appendix~\ref{NM_rev}, Einstein's equations in the Jordan
frame can be recast into the standard form if we define the effective
energy-momentum tensor given in \eqref{TEff}.  We then choose to define
$T^{(\phi, \rm eff)}_{\mu\nu}$ such that 
\begin{equation}
 T^{({\rm eff})}_{\mu\nu} = T^{(\phi, \rm eff)}_{\mu\nu} + \frac{M_{\rm Pl}^2}{f}T_{\mu\nu}^{(m)},
\end{equation} 
 i.e. we have
\begin{equation}
 T^{(\phi, \rm eff)}_{\mu\nu} = \frac{M_{\rm Pl}^2}{f}\left[T^{(\phi)}_{\mu\nu}+\nabla_\mu\nabla_\nu
	       f-g_{\mu\nu}\Box
	       f\right].
\end{equation}
As such, we see that despite the fact that $T^{(m)}_{\mu\nu}$ is
covariantly conserved, $T^{(\phi, \rm eff)}_{\mu\nu}$ is not, with 
\begin{equation}
 \nabla_\mu T^{(\phi, {\rm eff})\mu}{}_{\nu} = \frac{M_{\rm
  Pl}^2}{f^2}T^{(m)\mu}{}_\nu\nabla_\mu f.
\end{equation}
In a FLRW background we explicitly have
\begin{align}
 \rho_\phi^{\rm eff} &= \frac{M_{\rm
  Pl}^2}{f}\left[\frac{1}{2}h_{ab}\dot\phi^a\dot\phi^b + V -3H\dot
	    f\right] = \frac{f}{M_{\rm Pl}^2}\tilde\rho_\phi - 3\tilde
  H\frac{df}{d\tilde t}+\frac{3}{4f}\left(\frac{df}{d\tilde
 t}\right)^2,\\
p_\phi^{\rm eff} &= \frac{M_{\rm
  Pl}^2}{f}\left[\frac{1}{2}h_{ab}\dot\phi^a\dot\phi^b - V + \ddot f +2H\dot
	    f\right] = \frac{f}{M_{\rm Pl}^2}\tilde p_\phi
 +\frac{d^2f}{d\tilde t^2} + 2\tilde
  H\frac{df}{d\tilde t}-\frac{5}{4f}\left(\frac{df}{d\tilde
 t}\right)^2, 
\end{align}
and the continuity equation
\begin{equation}
 \dot\rho_\phi^{\rm eff} + 3H(\rho_\phi^{\rm eff} + p_\phi^{\rm eff}) =
  \frac{M_{\rm Pl}^2}{f^2}\dot f\rho^{(m)},
\end{equation}
where $\rho^{\rm eff}_\phi = - T^{(\phi, {\rm eff})\,0}{}_{0}$, $p^{\rm
eff}_\phi = T^{(\phi, {\rm eff})\,i}{}_{i}/3$, $\tilde\rho_\phi =
-\tilde T^{(\phi)0}{}_0$, $\tilde p_\phi = \tilde T^{(\phi)i}{}_i/3$ and
$\rho^{(m)} = -T^{(m)0}{}_0$.  The physical interpretation of this last
equation is less clear than that of \eqref{alphaCont} in the Einstein
frame.  However, one can assume that reheating completes when
$\rho^{(m)}\approx 3 (\rho_\phi^{\rm eff} + p_\phi^{\rm eff})/2$, where
we have used the fact that $\dot f/f\approx - 2H$, as can be seen from
\eqref{backRel}.

In the above analysis we have derived conditions for instant reheating
in both the Jordan and Einstein frames.  However, we note that imposing
instant reheating in this class of models gives rise to issues regarding
the discontinuity of $H$ or $\tilde H$, which results from the
assumption that $f\rightarrow M_{{\rm Pl}}^2$ instantaneously at the
time of reheating \cite{White:2013ufa}.  To avoid this issue one must
therefore solve the continuity equations dynamically.

\section{Summary and conclusions}\label{sum}

The high-precision nature of current CMB data dictates that reheating
dynamics must be taken into account when trying to constrain different
models of inflation.  Given the recent interest in inflation models
containing a non-minimal coupling to gravity and potentially multiple
scalar fields, in this paper we have revisited the process of
gravitational reheating that is inherent to this class of model.  Our
formulation allows for multiple, non-minimally coupled inflaton fields
endowed with a non-flat field-space metric, and it is assumed that these
fields are not directly coupled to matter.

At the level of the background dynamics, we saw that the oscillation of
the inflaton fields about their vacuum expectation values gives rise to
matter-dominated-like evolution of the Hubble rate in the Einstein
frame, as in elementary reheating scenarios.  In the Jordan frame, however, this
matter-dominated-like evolution is modulated by an oscillatory
component, and it is this oscillatory part that gives rise to the
gravitational particle production of minimally-coupled matter,
i.e. gravitational reheating.  When interpreted in the Einstein frame
the gravitational reheating does not result from the oscillatory nature
of the Hubble rate, but instead from the explicit interaction terms
between the inflaton sector and ordinary matter that are induced by
the conformal transformation.

In order to calculate the rate of particle production we used the method
of QFT in a classical background, which requires the calculation of
Bogoliubov coefficients.  Although this was not entirely necessary for
the perturbative reheating regime considered, the advantage is that much
of the discussion will also carry over to the resonant preheating regime, where the
perturbative flat-space QFT calculations are no longer applicable.
Taking appropriate limits, we were able to confirm agreement between the
Bogoliubov and perturbative QFT approaches, including kinematic
suppression factors.  Despite the difference in interpretation between
the Jordan and Einstein frames, we saw that the calculation of the
Bogoliubov coefficients associated with particle production was
independent of the frame in which we started.  This resulted from the
fact that the canonically normalised quantum fields one naturally
defines in the two frames are identical.

To finish, let us mention one possible extension of the framework
developed here.  In analysing the dynamics of the oscillating inflaton
fields at the end of inflation we made use of the mass eigen-basis of
the Einstein frame potential.  We implicitly made the assumption that
all of the fields begin oscillating about their vacuum expectation
values at approximately the same time, with $m_A\sim m_B\gg \tilde H$
for all $A$ and $B$.  Such an approximation, however, may not be valid.
Generally we might expect there to be a wide range of field masses, and
that different fields therefore begin to oscillate and decay at
different times.  In the case that heavier fields are present, which
start to oscillate and decay much earlier, it is perhaps reasonable to
assume that the resulting decay products are diluted by inflation --
which continues to be driven by the lighter fields -- and are therefore
negligible.  However, in the case that lighter spectator fields are
present, which do not oscillate and decay until much later, we might
expect them to play a significant role.  In general we would expect the
field space metric $h_{ab}$, potential $V(\phi)$ and non-minimal
coupling function $f(\phi)$ to all depend on these spectator fields.
Consequently, quantities such as $S_{ab}|_{\rm vev}$ and $\tilde
V_{ab}|_{\rm vev}$, which we took to be constants in the analysis of
Sec.~\ref{backdyn}, would all become functions of the spectator fields.
Ultimately, this would then lead to a spectator-field dependence of
$\tilde\Gamma_{\alpha^A\rightarrow\chi\chi}$ and
$\tilde\Gamma_{\alpha^A\rightarrow\overline\psi\psi}$, through their
dependence on $m_A$ and $f_A$, which would in turn give rise to a
modulated reheating scenario.  We leave further consideration of this
scenario to future work.

\acknowledgments This work was supported by Japan Society for the
Promotion of Science (JSPS) Research Fellowship for Young Scientists
No.~269337 (Y.W.) and JSPS Grant-in-Aid for Scientific Research (B)
No.~23340058 (J.W.). Y.W. acknowledges support from the Munich Institute
for Astro- and Particle Physics (MIAPP) of the Deutsche
Forschungsgemeinschaft (DFG) cluster of excellence ``Origin and
Structure of the Universe."

\appendix

\section{Energy-momentum tensors and their (non-)conservation}\label{NM_rev}

In this appendix we review in more detail the relation between
energy-momentum tensors defined in the Jordan and Einstein frames,
including determining whether or not they are covariantly conserved.  We
follow closely \cite{Maeda:1988ab,Fujii:2003dh,Koivisto:2005yk}, simply
generalising to the multi-field case.

\subsubsection*{Covariant conservation of the matter energy-momentum tensor in the Jordan frame}

Let us start by showing that the energy-momentum tensor for matter in
the Jordan frame is covariantly conserved, despite the presence of the
non-minimal coupling.  In Sec.~\ref{EEEM} we derived the
Einstein equations in the Jordan frame as
\begin{equation}\label{eq:1}
 G_{\mu\nu}=
  \frac{1}{f}\left[T^{(\phi)}_{\mu\nu}+T^{(m)}_{\mu\nu}+\nabla_\mu\nabla_\nu
	       f-g_{\mu\nu}\Box
	       f\right],
\end{equation}
where, using the definition
\begin{equation}
 T^{(i)}_{\mu\nu}=
  -\frac{2}{\sqrt{-g}}\frac{\delta\left(\sqrt{-g}\mathcal{L}^{(i)}\right)}{\delta
  g^{\mu\nu}},
\end{equation}
we have
\begin{equation}\label{eq:Tphi}
 T^{(\phi)}_{\mu\nu} = h_{ab}\nabla_\mu\phi^a\nabla_\nu\phi^b -
  g_{\mu\nu}\left(\frac{1}{2}h_{ab}g^{\rho\sigma}\nabla_\rho\phi^a\nabla_\sigma\phi^b
	    + V\right).
\end{equation}
We also gave the equations of motion for the fields as
\begin{equation}\label{eq:3}
 h_{ab}\Box\phi^b +
  \Gamma_{bc|a}g^{\mu\nu}\nabla_\mu\phi^b\nabla_\nu\phi^c - V_a + f_a R
  = 0,
\end{equation}
 where $\Gamma_{ab|c}=h_{cd}\Gamma^d_{ab}$ and $\Gamma^a_{bc}$ is the
 Christoffel connection associated with the field-space metric $h_{ab}$.
 Taking the covariant divergence of \eqref{eq:1}, and using the Bianchi
 identity, we find
\begin{equation}\label{eq:EED}
 \nabla^\mu G_{\mu\nu}\equiv 0=-\frac{\nabla^\mu f}{f}G_{\mu\nu} +
  \frac{1}{f}\left[\nabla^\mu T^{(\phi)}_{\mu\nu}+\nabla^\mu T^{(m)}_{\mu\nu}+\Box\nabla_\nu
	       f-\nabla_\nu\Box
	       f\right].
\end{equation}
Similarly, taking the covariant derivative of \eqref{eq:Tphi} and using the equations
of motion \eqref{eq:3}, we also find
\begin{equation}
 \nabla^\mu T^{(\phi)}_{\mu\nu} = -R \nabla_\nu f.
\end{equation}
Substituting this result into \eqref{eq:EED}, and recalling
$G_{\mu\nu}=R_{\mu\nu}-\frac{1}{2}g_{\mu\nu}R$,
we arrive at
\begin{equation}
 \nabla^\mu G_{\mu\nu}\equiv 0 = -\frac{\nabla^\mu
  f}{f}R_{\mu\nu}+\frac{1}{f}\left[\nabla^\mu T^{(m)}_{\mu\nu} +
			       \Box\nabla_\nu f- \nabla_\nu\Box f\right].
\end{equation}
However, from the definition of the Riemann tensor we have
\begin{equation}
 \Box\nabla_\nu f - \nabla_\nu\Box f = R_{\nu\mu}\nabla^\mu f,
\end{equation}
which clearly then leaves us with 
\begin{equation}
  \nabla^\mu G_{\mu\nu}\equiv 0 =\frac{1}{f}\nabla^\mu T^{(m)}_{\mu\nu},
\end{equation}
i.e. we have recovered the fact that the matter energy-momentum tensor
in the Jordan frame is covariantly conserved.     

One could also consider
an effective energy-momentum tensor defined by \eqref{eq:1} as
\begin{equation}\label{TEff}
 T_{\mu\nu}^{({\rm eff})}=\frac{M_{\rm Pl}^2}{f}\left[T^{(\phi)}_{\mu\nu}+T^{(m)}_{\mu\nu}+\nabla_\mu\nabla_\nu
	       f-g_{\mu\nu}\Box
	       f\right],
\end{equation} 
such that Einstein's equations take the standard form $G_{\mu\nu} =
T^{({\rm eff})}_{\mu\nu}/M_{\rm Pl}^2$.  This effective energy-momentum tensor is of course
covariantly conserved as a result of the Bianchi identity.

\subsubsection*{Energy-momentum tensors in the Einstein frame}

Turning to the Einstein frame, if we assume that $\mathcal{L}^{(m)}$ only depends on $g_{\mu\nu}$
and not its derivatives, then we can write
\begin{equation}\label{eq:ETJ}
 T^{(m)}_{\mu\nu}=-\frac{2}{\sqrt{-g}}\frac{\delta\left(\sqrt{-g}\mathcal{L}^{(m)}\right)}{\delta
  g^{\mu\nu}}=-\frac{2}{\sqrt{-g}}\frac{\delta\left(\sqrt{-g}\mathcal{L}^{(m)}\right)}{\delta
  \tilde{g}^{\rho\sigma}}\frac{\partial\tilde{g}^{\rho\sigma}}{\partial g^{\mu\nu}}.
\end{equation} 
Under the conformal transformation we have
$\tilde{g}_{\mu\nu}=\Omega^2 g_{\mu\nu}$,
$\tilde{g}^{\mu\nu}=g^{\mu\nu}/\Omega^2$ and
$\sqrt{-\tilde{g}}=\Omega^4\sqrt{-g}$, which we can substitute into
\eqref{eq:ETJ} to find 
\begin{equation}\label{JFTTEFT}
 T^{(m)}_{\mu\nu}=\Omega^2\tilde{T}^{(m)}_{\mu\nu}.
\end{equation} 
Recall that $\Omega^2 = f(\bm \phi)/M_{\rm Pl}^2$.  Taking the covariant
divergence of this (with respect to the Jordan frame metric) we have
\begin{equation}\label{eq:DJEMT}
 \nabla^\mu T^{(m)}_{\mu\nu} =
  \Omega^4\tilde{g}^{\mu\alpha}\nabla_\alpha\tilde{T}^{(m)}_{\mu\nu}
  + 2\Omega^3\tilde{g}^{\mu\alpha}\Omega_{\alpha}\tilde{T}^{(m)}_{\mu\nu}.
\end{equation}
We now need to use the relation between covariant derivatives as defined
with respect to $g_{\mu\nu}$ and $\tilde{g}_{\mu\nu}$.  The relation
between the Christoffel symbols is given as
\begin{equation}
 \Gamma^\alpha_{\beta\gamma} =
  \tilde{\Gamma}^\alpha_{\beta\gamma}-\frac{1}{\Omega}\left(\delta^\alpha_\beta\Omega_{\gamma}+\delta^\alpha_\gamma\Omega_{\beta}-
						       \tilde{g}^{\alpha\rho}\tilde{g}_{\gamma\beta}\Omega_{\rho}\right),
\end{equation}
and on substituting this result into \eqref{eq:DJEMT} we find
\begin{equation}\label{EMTCEF}
 \nabla^\mu T^{(m)}_{\mu\nu} =0= \Omega^4\tilde{\nabla}^\mu
  \tilde{T}^{(m)}_{\mu\nu} + \tilde{T}^{(m)}\Omega^3\Omega_{\nu}\qquad\Rightarrow\qquad\tilde{\nabla}^\mu
  \tilde{T}^{(m)}_{\mu\nu} = -\tilde{T}^{(m)}\frac{\Omega_{\nu}}{\Omega},
\end{equation}
where $\tilde{T}^{(m)}$ is the trace of the matter energy-momentum
tensor.  Thus, we see that even if $T^{(m)}_{\mu\nu}$ is covariantly
conserved, in general $\tilde{T}^{(m)}_{\mu\nu}$ is not.  It will,
however, be conserved if $\tilde{T}^{(m)}=0$, which is the case for
radiation-like matter.

Given that in the Einstein frame we have 
\begin{equation}\label{eq:EFC}
 \tilde{\nabla}^\mu\tilde{G}_{\mu\nu}\equiv 0 = \tilde{\nabla}^\mu\left(\tilde{T}^{(\phi)}_{\mu\nu}+\tilde{T}^{(m)}_{\mu\nu}\right),
\end{equation}   
the non-conservation of $\tilde{T}^{(m)}_{\mu\nu}$ implies a
non-conservation of $\tilde{T}^{(\phi)}_{\mu\nu}$.  Let us try to
determine this explicitly.

When we try to calculate the equations of motion for the scalar fields
in the Einstein frame, we need to correctly take into account the
dependence of $\sqrt{-g}\mathcal{L}^{(m)}$ on $\phi^a$ that results from
the conformal transformation.  However, as the only dependence of
$\sqrt{-g}\mathcal{L}^{(m)}$ on $\phi^a$ comes from the conformal
transformation $\tilde{g}_{\mu\nu}= \Omega^2 g_{\mu\nu}$, if we once
again assume that $\sqrt{-g}\mathcal{L}^{(m)}$ only depends on
$g_{\mu\nu}$ and not its derivatives, then we can simply use the rules
of partial differentiation to obtain
\begin{equation}
 \frac{\delta\left(\sqrt{-g}\mathcal{L}^{(m)}\right)}{\delta\phi^a} =
  \frac{\delta\left(\sqrt{-g}\mathcal{L}^{(m)}\right)}{\delta
  g^{\mu\nu}}\frac{\partial g^{\mu\nu}}{\partial \phi^a} = 2\Omega\Omega_{a}\tilde{g}^{\mu\nu}\frac{\delta\left(\sqrt{-g}\mathcal{L}^{(m)}\right)}{\delta
  g^{\mu\nu}}=2\frac{\Omega_{a}}{\Omega}\tilde{g}^{\mu\nu}\frac{\delta\left(\sqrt{-g}\mathcal{L}^{(m)}\right)}{\delta
  \tilde{g}^{\mu\nu}}=-\frac{\Omega_{a}}{\Omega}\sqrt{-\tilde{g}}\tilde{T}^{(m)}.
\end{equation} 
As such, the equations of motion for the scalar fields become
\begin{equation}\label{eq:7}
 -S_{ab}\tilde{\Box}\phi^b -
  ^{(S)}\Gamma_{bc|a}\tilde{g}^{\mu\nu}\tilde{\nabla}_\mu\phi^b\tilde{\nabla}_\nu\phi^c
  + \tilde{V}_{,a} + \frac{\Omega_{a}}{\Omega}\tilde{T}^{(m)} = 0,
\end{equation}
where ${}^{(S)}\Gamma_{bc|a}=S_{ad}{}^{(S)}\Gamma^d_{bc}$ and
${}^{(S)}\Gamma^d_{bc}$ is the Christoffel connection associated with $S_{ab}$. 
As given in the main text, the energy momentum tensor
$\tilde{T}^{(\phi)}_{\mu\nu}$ takes the form 
\begin{equation}
 \tilde{T}^{(\phi)}_{\mu\nu}=S_{ab}\tilde{\nabla}_\mu\phi^a\tilde{\nabla}_\nu\phi^b
  -
  \tilde{g}_{\mu\nu}\left(\frac{1}{2}S_{ab}\tilde{g}^{\rho\sigma}\tilde{\nabla}_\rho\phi^a\tilde{\nabla}_\sigma\phi^b
		    + \tilde{V}\right).
\end{equation}
Taking the covariant divergence of this and making use of the equations
of motion \eqref{eq:7}, we find
\begin{equation}
 \tilde{\nabla}^\mu\tilde{T}_{\mu\nu}^{(\phi)} = \frac{\Omega_{\nu}}{\Omega}\tilde{T}^{(m)}.
\end{equation}
This is entirely consistent with what we expected from \eqref{eq:EFC}
and \eqref{EMTCEF}.  

\section{Details regarding fermion particle production}\label{fermApp}

In this appendix we give additional details regarding the fermion
particle production calculation.  As stated in the main text, we follow
closely the analyses of
\cite{Parker:1971pt,Peloso:2000hy,Nilles:2001fg,Chung:2011ck}.

\subsubsection*{Conventions}

First let us clarify our conventions.  We use the $(-+++)$ sign
convention and the Dirac gamma matrix representation
\begin{equation}
\gamma^0 = \left(\begin{array}{cc}-i& 0\\ 0&i\end{array}\right),\quad\gamma^i=\left(\begin{array}{cc}0&-i\sigma^i\\ i\sigma^i&0\end{array}\right),
\end{equation}
where $\sigma^i$ are the Pauli matrices.

Following \cite{Chung:2011ck}, for the eigenvectors of the helicity operator 
we use the following spherical-coordinates-based representation 
\begin{equation}
h_{+1}(\hat k)=\left(
 \begin{array}{c}
  \cos\frac{\theta_{\hat k}}{2}e^{-i\phi_{\hat k}}\\\sin\frac{\theta_{\hat k}}{2}
 \end{array}\right),\qquad
h_{-1}(\hat k)=\left(
 \begin{array}{c}
  \sin\frac{\theta_{\hat k}}{2}e^{-i\phi_{\hat k}}\\ -\cos\frac{\theta_{\hat k}}{2}
 \end{array}\right),
\end{equation}
where we have chosen the normalisation such that $h_r^\dagger(\hat
k)h_s(\hat k)=\delta_{rs}$ is satisfied.  With this choice, it is then
possible to show that
\begin{equation}
 -i\sigma^2 h_r^\ast(\hat k) = -re^{i\phi_{\hat k}} h_{-r}(\hat{k}),
\end{equation}
which on using $h_r(-\hat k)=h_{-r}(\hat k)$ allows us to find 
\begin{equation}
 V_r(\vec k)=\gamma^2U^\ast_r({\vec k},x)=\frac{e^{i\phi_{\hat k}}}{(2\pi)^{3/2}}\left(\begin{array}{c}
				       -u^\ast_{\mathcal B}(k,\eta)h_r(-\hat{k})\\
					     ru_{\mathcal A}^\ast(k,\eta) h_r(-\hat{k})
					    \end{array}\right)e^{-i\vec
 k\cdot \vec x}.
\end{equation}

\subsubsection*{Calculation of production rate}

Here we give more details regarding the calculation of the production
rate for fermions.  As with the scalar case, in order to determine
the particle production rate we need to evaluate the right-hand side of
the continuity equation for the fermion energy-momentum tensor in the
Einstein frame.  As such, we need to first determine the vacuum
expectation values of the components of the energy-
momentum tensor, which we denote $\langle 0|\tilde
T^{(\tilde\psi)\mu}{}_\nu|0\rangle = {\rm diag}(-\tilde\rho_\psi,~\tilde
p_\psi,~\tilde p_\psi,~\tilde p_\psi)$.  In terms of the quantities
$\tilde\rho_\psi$ and $\tilde p_\psi$ the continuity equation takes the
form
\begin{equation}
 \frac{1}{\tilde a^4}\frac{d}{d\tilde t}(\tilde a^4 \tilde\rho_\psi) +
  \tilde H(-\tilde \rho_\psi + 3\tilde p_\psi) =
  \frac{1}{2f}\frac{df}{d\tilde t}(-\tilde \rho_\psi + 3 \tilde p_\psi).
\end{equation}
The explicit form of $\tilde T^{(\tilde\psi)\mu}{}_\nu$ for fermions is
given in \eqref{fermTmunu}, and on writing in terms of the canonically
normalised field $\Psi(x)$ given in \eqref{genspin} and taking the
vacuum expectation value we find
\begin{align}
 \tilde\rho_\psi &= \frac{4}{(2\pi)^3\tilde a^4}\int d^3k
  w_k\left(|\mathcal B_k(\eta)|^2-\frac{1}{2}\right),\\
-\tilde\rho_\psi + 3\tilde p_\psi &= \frac{4\tilde a m_\psi}{\tilde
 a^4\Omega}\int \frac{d^3k}{(2\pi)^3}\left[\frac{\tilde a
 m_\psi}{w_k\Omega}\left(\frac{1}{2}-|\mathcal B_k(\eta)|^2\right)-\frac{k}{w_k}\Re\left(\mathcal
 A_k(\eta)\mathcal B_k^\ast(\eta) e^{-2i\int w_kd\eta^\prime}\right)\right].
\end{align}
If we are assuming that initially no fermion particles are
present, then it is appropriate to consider the perturbative regime where $\mathcal
B_k(\eta)\ll 1$ and $\mathcal A_k(\eta)\simeq 1$.  As such, to leading
order in $\mathcal B_k(\eta)$ and $\alpha^A$ we find
\begin{equation}\label{fermPP}
 \frac{1}{2f}\frac{df}{d\tilde t}(-\tilde\rho_\psi+3\tilde p_\psi) \simeq
  -\frac{1}{2f}\frac{df}{d\tilde t}\frac{4(4\pi)}{(2\pi)^3\tilde
  a^4}\int dk k^2 \frac{k\tilde a m_\psi}{w_k}\Re\left(\mathcal
						  B_k^\ast(\eta)e^{-2i\int
						  w_kd\eta^\prime}\right).
\end{equation}

It is now clear we must solve for $\mathcal B_k(\eta)$, i.e. solve
Eq. \eqref{fermBog2}.  In order to do so, we note that to leading
order in $\alpha^A$ we have
\begin{align}\label{fermFreq}
  w_k^2 &\simeq k^2 + (\tilde{a}m_\psi)^2,\\\label{fermFP}
 (am_\psi)^\prime &\simeq
  -\tilde{a}m_\psi\frac{f_A\alpha^{A\prime}}{2M_{\rm Pl}^2} + \tilde
 a^2\tilde H m_\psi.
 \end{align}
Note that, as with the bosonic case, taking expressions to leading order
in $\alpha^A$ ensures that we are only considering the tri-linear
interaction terms and the perturbative regime appropriate for comparison
with the perturbative QFT calculations of Sec.~\ref{QFT}.  Taking
$\mathcal A_k(\eta)\rightarrow 1$ and integrating \eqref{fermBog2}
we have
\begin{equation}
 \mathcal B_k(\eta) = \int_{\eta_0}^\eta d\eta^\prime
  \frac{km_\psi}{2w_k^2}\left(\tilde a^2\tilde H^2 -\tilde
			 a\frac{f_A\alpha^{A\prime}}{2M_{\rm Pl}^2}\right)e^{-2i\int^{\eta^\prime}_{-\infty}w_kd\eta^{\prime\prime}}.
\end{equation}
The first term in the brackets is slowly varying, so that its
contribution to $\mathcal B_k(\eta)$ averages to zero.  The second term, however,
is highly oscillatory, thus giving a non-zero contribution to $\mathcal
B_k(\eta)$ that can be calculated using the stationary phase
approximation.  Explicitly, we have
\begin{equation}
 \mathcal B_k(\eta) = \sum_A\int_{\eta_0}^\eta d\eta^\prime
  \frac{k\tilde a^{1/2} f_A m_A\alpha_0^Am_\psi}{8M_{\rm Pl}^2iw_k^2}\left(e^{im_A\psi_{k,1}^A(\eta^\prime)}-e^{im_A\psi_{k,1}^A(\eta^\prime)}\right),
\end{equation}
where $\psi_{k,1}^A$ and $\psi_{k,2}^A$ are as given in \eqref{phase1}
and \eqref{phase2} but with $w_k^2 = k^2 + a^2m_\psi^2$.  The phase
$\psi_{k,2}^A$ has no stationary points for physical values of $\tilde
a$, and therefore the second term in the above expression gives no
contribution to $\mathcal B_k(\eta)$.  The first term involving the
phase $\psi_{k,1}^A$, however, does have a stationary point at $\tilde a
= 2w_k/m_A$.  To leading order in $\alpha^A$, this stationary phase
condition gives
\begin{equation}\label{eq2}
 \frac{k}{\tilde{a}(\eta^A_k)} \simeq \frac{m_A}{2}\left(1-\frac{4m_\psi^2}{m_{\hat A}^2}\right)^{1/2},
\end{equation}
where $\eta_k^A$ denotes the time at which the condition is satisfied
for a given $k$ and $m_A$.  As in the scalar case, this coincides with
our expectation from kinematics.  In making the stationary phase
approximation one also needs the second derivative of the phase at the
time $\eta_k^A$, which in the fermionic case is given as
\begin{equation}\label{fermPsiDeriv}
\psi_{k,1}^{A\prime\prime}(\eta_k^{\hat A}) \simeq \tilde a^2(\eta_k^{\hat A})\tilde
 H(\eta_k^{\hat A})\left(1-\frac{4m_\psi^2}{m_A^2}\right) + \sum_B2\tilde
 a(\eta_k^{\hat A})\frac{m_\psi^2}{m_A^2}\frac{f_B\alpha^{B\prime}(\eta_k^{\hat A})}{M_{\rm Pl}^2}.
\end{equation}  
The final solution for $\mathcal B_k(\eta)$ is given in \eqref{fermBeta}.

We next turn to evaluating \eqref{fermPP}.  On expanding $df/d\tilde t$
we obtain
\begin{equation}
 \frac{1}{2f}\frac{df}{d\tilde t}(-\tilde\rho_\psi+3\tilde p_\psi) =
  \sum_A\frac{4(4\pi)}{(2\pi)^3\tilde a^5}\int dk k^2w_k\frac{k\tilde
  a^{1/2}f_A\alpha_0^Am_Am_\psi}{8M_{\rm Pl}^2iw_k^2}2i\Im\left[\mathcal
							   B_k^\ast(\eta)\left(e^{im_A\psi_{k,1}^A(\eta)}-e^{im_A\psi_{k,2}^A(\eta)}\right)\right].
\end{equation}
We see that this is a highly oscillatory function, and as with the
scalar case we consider taking an average over several oscillations as 
\begin{align}
 \left\langle\frac{1}{2f}\frac{df}{d\tilde t}(-\tilde\rho_\psi+3\tilde p_\psi)\right\rangle &=
  \frac{1}{2T}\int^{\eta + T}_{\eta - T}d\eta^\prime\sum_A\frac{4(4\pi)}{(2\pi)^3\tilde a^5}\int dk k^2w_k\frac{k\tilde
  a^{1/2}f_A\alpha_0^Am_Am_\psi}{8M_{\rm
  Pl}^2iw_k^2}\\\nonumber&\hspace{5cm}\times 2i\Im\left[\mathcal
							   B_k^\ast(\eta^\prime)\left(e^{im_A\psi_{k,1}^A(\eta^\prime)}-e^{im_A\psi_{k,2}^A(\eta^\prime)}\right)\right],
\end{align}
where $T\sim \mathcal O(1/(\tilde a m_A))$.  This averaged quantity will
only be non-zero if $\eta$ coincides with a stationary point of the
phases $\psi_{k,1}^A(\eta)$ or $\psi_{k,2}^A(\eta)$.  Seeing as
$\psi_{k,2}^A(\eta)$ has no stationary points for physical values of
$\tilde a$, we only obtain contributions from terms involving
$\psi_{k,1}^A(\eta)$.  On making the stationary phase approximation we
arrive at
\begin{align}
 \left\langle\frac{1}{2f}\frac{df}{d\tilde t}(-\tilde\rho_\psi+3\tilde
  p_\psi)\right\rangle &= \sum_{A,B}\frac{4(4\pi)}{(2\pi)^3\tilde
 a^5(\eta)}\int dk
 \delta(\eta - \eta_k^A)w_k(\eta_k^A) k^2\\\nonumber &\hspace{4cm}\times 2\Re\left[\mathcal B_k^A\mathcal B_k^{B\ast}e^{im_A\psi_{k,1}^A(\eta_k^A) -
	  im_B\psi_{k,1}^B(\eta_k^B) + i(s_k^A-s_k^B)\pi/4}\right]\\\nonumber&\hspace{4cm}\times\Theta(\eta_k^A-\eta_k^B)\Theta(\eta_k^B-\eta_0),
\end{align}  
where $\mathcal B_k^A$ is as defined in \eqref{BkA}.  On using the fact
that $\mathcal B_k^A\mathcal B_k^{B\ast} = \mathcal B_k^{A\ast}B_k^B$,
this can then be written as
\begin{align}\label{fermTotal}
 \left\langle\frac{1}{2f}\frac{df}{d\tilde t}(-\tilde\rho_\psi+3\tilde
  p_\psi)\right\rangle &= 
\sum_{A}\frac{4(4\pi)}{(2\pi)^3\tilde
 a^5(\eta)}\int dk
 \delta(\eta - \eta_k^A)w_k(\eta_k^A) k^2|\mathcal B_k^A|^2\\\nonumber&\quad+\sum_{A,B>A}\frac{4(4\pi)}{(2\pi)^3\tilde
 a^5(\eta)}\int dk
 \delta(\eta - \eta_k^A)w_k(\eta_k^A) k^2\mathcal B_k^A\mathcal B_k^{B\ast}\\\nonumber&\hspace{2cm}\times 2\cos\left(m_A\psi_{k,1}^A(\eta_k^A) -
	  m_B\psi_{k,1}^B(\eta_k^B) + (s_k^A-s_k^B)\pi/4\right)\Theta(\eta_k^B-\eta_0).
\end{align}
In the above expression we have arranged that $m_B>m_A$ for $B>A$,
meaning that $\Theta(\eta_k^A-\eta_k^B)=1$ only for $B>A$.  Next we re-write the
delta function in $\eta$ as a delta function in $k$.  The relation is as
given in \eqref{deltaConv}, but now with $\psi_{k,1}^{A\prime\prime}(\eta_k^A)$ as
given in \eqref{fermPsiDeriv} and $\mu_A$ to leading order in $\alpha^A$
given as
\begin{equation}
 \mu_A \simeq \frac{m_A}{2}\left(1-\frac{4m_\psi^2}{m_{\hat A}^2}\right)^{1/2}.
\end{equation}   
Assuming that the non-diagonal terms in the second line of \eqref{fermTotal} average to zero, the diagonal terms give rise to \eqref{fermResult}.  
\bibliography{bibliography}{}

\end{document}